\documentclass[twoside,slac_one]{revtex4}
\usepackage{graphicx}
\usepackage{fancyhdr}
\usepackage{amsmath} 
\usepackage{bm}
\usepackage{amsxtra}
\usepackage{amssymb}
\usepackage{amsthm}
\usepackage{latexsym}
\usepackage{lscape}

\begin{document}

\title{Measurements of {\boldmath $CP$} Violation in {\boldmath $B$} Meson decays at Belle}

%

\author{Himansu Bhusan Sahoo on behalf of the Belle Collaboration}
\affiliation{Department of Physics and Astronomy, University of Hawaii, Honolulu, HI 96822, USA\\
himansu@phys.hawaii.edu}

\begin{abstract}
In this proceeding, we report the recent measurements of time-dependent 
$CP$ violation in $B$ meson decays from the Belle Collaboration.
The Belle experiment stopped operation in June 2010 and collected
nearly $772 \times 10^{6}$ $B\overline{B}$ pairs 
at the $\Upsilon(4S)$ resonance at the KEKB 
asymmetric-energy $e^+e^-$ collider.
We used this full data sample to measure the $CP$ violation parameters in
$b \to c \bar{c} s$ and $b \to c \bar{c} d$ decays.
We also report new measurements of $CPT$ violation parameters 
in $B$ decays.
Furthermore, we report the first observation of a new $b \to s$ 
radiative penguin decay $B^0 \to \phi K^0 \gamma$, as well as
measurements of its time-dependent $CP$ asymmetry.
\end{abstract}

\maketitle

\thispagestyle{fancy}


\section{Introduction}
In the standard model (SM), $CP$ violation in $B^0$
meson decays originates from an irreducible complex phase 
in the $3 \times 3$ Cabibbo-Kobayashi-Maskawa (CKM) 
mixing matrix~\cite{ckm}.
The unitarity condition of the CKM matrix gives rise to a relation
$V_{ud}V_{ub}^*+V_{cd}V_{cb}^*+V_{td}V_{tb}^* = 0$, which can be represented by a triangle in the complex plane, known as the Unitarity Triangle (UT).
The main objective of the $B$-factories is to test the SM picture of the origin of $CP$ violation by measuring
 the angles (denoted by $\phi_1$, $\phi_2$ and $\phi_3$)\footnote{BaBar uses an alternative notation $\beta$, $\alpha$ and $\gamma$ corresponding to $\phi_1$, $\phi_2$ and $\phi_3$.} and sides of the UT using different $B$ decays. 
In this proceeding, we report the recent measurements of
time-dependent $CP$ asymmetries in $B \to$ Charmonium $K^0$ decays.
These proceed via $b \to c \overline{c} s$ tree diagram and 
provide a precise measurement of the angle $\phi_1$ ($\equiv \pi - \arg(V_{tb}^*V_{td}/V_{cb}^*V_{cd})$).
For such decays the interference between the tree amplitude 
and the amplitude from $B^0-\overline{B}{}^0$ mixing is dominated 
by the single phase $\phi_1$.
The other modes which allows measurements of $\phi_1$ are
$b \to c\overline{c}d$ transitions like 
$B^0 \to D^+ D^-$ and $B^0 \to D^{*+} D^{*-}$. 
These modes are dominated by tree diagrams, but loops may
contribute. So, sensitivity to new physics (NP) increases in these decays.

On the other hand, rare radiative $B$ meson decays are
flavor changing neutral currents,
forbidden at tree level in the SM but allowed through
electroweak loop processes. 
The loop can be mediated by non-SM particles
(for example, charged Higgs or SUSY particles)
and therefore is sensitive to NP.
The emitted photons 
in $b \rightarrow s \gamma$ 
($\overline{b} \rightarrow \overline{s} \gamma$) decays
are predominantly left-handed (right-handed) in the SM, and
hence the time-dependent $CP$ asymmetry 
is suppressed by the quark mass ratio ($2m_s/m_b$).
The expected mixing-induced $CP$ asymmetry parameter (${\cal S}$)
is ${\cal O}(3\%)$
and the direct $CP$ asymmetry parameter 
(${\cal A}$) is $\sim 0.6\%$~\cite{ags}.
In several extensions of the SM, both photon helicities can
contribute to the decay.
Therefore, any significantly larger $CP$ asymmetry 
would be clear evidence for NP.

\section{Experimental Apparatus}
The Belle detector is a large-solid-angle magnetic
spectrometer that consists of a silicon vertex detector (SVD),
a 50-layer central drift chamber (CDC), an array of
aerogel threshold Cherenkov counters (ACC),
a barrel-like arrangement of time-of-flight
scintillation counters (TOF), and an electromagnetic calorimeter (ECL)
comprised of CsI(Tl) crystals located inside
a superconducting solenoid coil that provides a 1.5~T
magnetic field.  An iron flux-return located outside 
the coil is instrumented to detect $K_L^0$ mesons and to identify
muons (KLM).  The detector is described in detail elsewhere~\cite{Belle}.
Two different inner detector configurations were used. 
For the first sample
of $152 \times 10^6$ $B\overline{B}$ pairs, a 2.0 cm radius beampipe
and a 3-layer silicon vertex detector (SVD1) were used;
for the latter samples,
a 1.5 cm radius beampipe, a 4-layer silicon detector (SVD2),
and a small-cell inner drift chamber were used.
Belle stopped operation in June 2010 and collected more than
710 ${\rm fb}^{-1}$ of data. This corresponds to a total of 
$772 \times 10^6$ $B\bar{B}$ pairs at $\Upsilon(4S)$ resonance.
Out of these nearly $80\%$ of the data sample is reprocessed with a new
tracking algorithm.

\section{Analysis Technique}

In the $B$ meson system, the $CP$ violating asymmetry lies 
in the time-dependent
decay rates of the $B^0$ and $\overline{B}{}^0$ decays to a 
common $CP$-eigenstate ($f_{CP}$). The asymmetry can be written as:
\begin{eqnarray*}
{\cal A}_{CP}(t) &=& \frac{\Gamma[\overline{B}{}^0(t)\to f_{CP}] -  \Gamma[B^0(t)\to f_{CP}]} 
                               {\Gamma[\overline{B}{}^0(t)\to f_{CP}] +  \Gamma[B^0(t)\to f_{CP}]}\\
                       &=& {\cal S} \sin (\Delta m_d t) + {\cal A} \cos(\Delta m_d t)
\end{eqnarray*}
where
\begin{equation}
\mathcal{S} = \frac{2 \,\rm{Im\lambda}}{|\lambda|^2+1} \hspace{2cm}
\mathcal{A} = \frac{|\lambda|^2-1}{|\lambda|^2+1}.
\end{equation}

\noindent Here $\Gamma(B^0 (\overline{B}{}^0) \to f_{CP})$ is the decay rate 
of a $B^0 (\overline{B}{}^0)$ meson decays to $f_{CP}$ 
at a proper time $t$ after the production,
$\Delta m_d$ is the mass difference 
between the two neutral $B$ mass eigenstates, $\lambda$ is a complex parameter
depending on the $B^0-\overline{B}{}^0$ mixing as well as the decay 
amplitudes of the $B$ meson decays to the $CP$ eigenstate.
The parameter ${\cal S}$ is the measure of 
mixing-induced $CP$ violation, whereas
${\cal A}$ is the measure of direct $CP$ violation\footnote{Note that BaBar uses the convention ${\cal C}$ = $-{\cal A}$.}.

In the $B$ factories, in order to measure the time-dependent 
$CP$ violation parameters, we fully
reconstruct one neutral $B$ meson in its decay into a $CP$ eigenstate.
From the remaining particles in the event, 
the vertex of the other $B$ meson is reconstructed 
and its flavor is identified.
In the decay chain 
$\Upsilon(4S)\to B^0 \overline{B}{}^0 \to f_{CP} f_{\rm tag}$, 
where one of the $B$ mesons decays at time $t_{CP}$ 
to a $CP$ eigenstate $f_{CP}$, which
is our signal mode, and the other decays at time $t_{\rm tag}$ 
to a final state $f_{\rm tag}$ that distinguishes between 
$B^0$ and $\overline{B}{}^0$, the decay
rate has a time dependence given by~\cite{carter}
\begin{eqnarray}
{\cal P}(\Delta{t})= \frac{ e^{-|\Delta{t}|/{\tau_{B^0}}} }{4\tau_{B^0}}
\biggl\{1 & + & q \cdot 
 \Bigl[ {\cal S} \sin(\Delta m_d \Delta{t}) 
 + {\cal A} \cos(\Delta m_d \Delta{t})
\Bigr] \biggr\}.
\label{eq_decay}
\end{eqnarray}
\noindent Here $\tau_{B^0}$ is the neutral $B$ lifetime,
$\Delta t = t_{CP} - t_{\rm{tag}}$,
and the $b$-flavor charge $q$ equals $+1$ ($-1$) 
when the tagging $B$ meson is identified as 
$B^0$ ($\overline{B}{}^0$).
Since the $B^0$ and $\overline{B}{}^0$ are approximately 
at rest in the $\Upsilon(4S)$ center-of-mass system,
$\Delta t$ can be determined from the displacement in $z$ 
between the $f_{CP}$ and $f_{\rm tag}$
decay vertices: $\Delta t \simeq \Delta z / (\beta \gamma c)$,
where $c$ is the speed of light.
The vertex position of the $f_{CP}$ decay is reconstructed using charged tracks
(for example, lepton tracks from $J/\psi$ in $B^0 \to J/\psi K_S^0$ decays)
and that of the $f_{\rm tag}$ decay from well-reconstructed tracks that are not
assigned to $f_{CP}$~\cite{vertexing}
The $\Delta z$ is approximately 200 $\mu$m in Belle and 250 $\mu$m in BaBar.
We also consider the effect of detector resolution and mis-identification
of the flavor~\cite{tagging}. 
We use a flavor tagging algorithm
to obtain the $b$-flavor charge $q$ and a
tagging quality factor $r \in \left[0,1\right]$.
The value $r=0$ signifies no flavor discrimination while $r=1$ implies
unambiguous flavor assignment. 
The data are divided into seven $r$ intervals.
Finally, the $CP$ violation parameters are obtained from an
unbinned maximum likelihood (UML) fit to the $\Delta t$ distribution.
%


\section{{\boldmath $b \to c\bar{c}s$} Decay Modes}

The $B \to$ Charmonium $K^0$ decays that are mediated by
$b \to c\overline{c}s$ decays are known as the golden modes for
$CP$ violation measurements.
They have clean experimental signatures: many accessible modes
with relatively large branching fractions $\mathcal{O}(10^{-4})$, 
low experimental background levels and 
high reconstruction efficiencies.
These modes are dominated by
a color-suppressed $b \to c\overline{c} s$ tree diagram and the dominant
penguin diagram has the same weak phase.
The $CP$ violation comes from the $V_{td}$ element in the mixing box 
diagram, which contains the phase.
For $f_{CP}$ final states resulting from a $b \to c\bar{c}s$ transition, 
the SM predicts
$\mathcal{S} = - \xi_{CP}\sin 2 \phi_1$ and $\mathcal{A} = 0$,
where $\xi_{CP}$ is the $CP$ eigenvalue of the final state and is
$+1$($-1$) for $CP$-even ($CP$-odd) states.
The asymmetry is given as
\begin{equation}
{\cal A}_{CP} = -\xi_{CP} \sin(2\phi_1) \sin(\Delta m \Delta t).
\end{equation}
We can verify this experimentally 
by measuring the number of $B^0 (\overline{B}{}^0)$ 
decays to $CP$ eigenstates.
A non-zero value of ${\cal A}$ or any measurement of $\sin 2\phi_1$ 
that has a significant deviation indicates evidence for NP.

Belle recently reported new measurements with its full data sample
using the modes
$B^0 \to J/\psi K^0$, $B^0 \to \psi' K_S^0$ and $B^0 \to \chi_{c1} K_S^0$.
The $J/\psi$ candidates are reconstructed from their decays to
$e^+e^-$ and $\mu^+\mu^-$, with the $K_S^0$ reconstructed from $\pi^+\pi^-$.
The $\psi'$ candidates are reconstructed from $e^+e^-$, $\mu^+\mu^-$ as well as
$J/\psi \pi^+ \pi^-$ decays. 
The $\chi_{c1}$ is reconstructed from its decays to $J/\psi \gamma$.
$B$ candidates are identified using 
two kinematic variables: the energy difference 
$\Delta E \equiv E_B^{\rm cms} - E_{\rm beam}^{\rm cms}$ and the
beam-energy-constrained mass 
$M_{\rm bc} \equiv \sqrt{(E_{\rm beam}^{\rm cms}/c^2)^2 - (p_B^{\rm cms}/c)^2}$,
where $E_{\rm beam}^{\rm cms}$ is the beam energy in the cms, and 
$E_B^{\rm cms}$ and $p_B^{\rm cms}$ are the center-of-mass (cms) 
energy and momentum, respectively, of the reconstructed $B$ candidate.
Belle reported nearly 15600 $CP$-odd signal events with a purity of $96\%$
and nearly 10000 $CP$-even signal events with a purity of $63\%$.
Belle observed $CP$ violation in all the listed charmonium modes and 
the results are summarized in Table~\ref{tab:charmoniumresult}.

\begin{table}[hbtp]
\begin{center}
  \caption{The $CP$-violating parameters measured by Belle with golden modes 
    using a data sample (the errors are statistical only). 
    Belle observed $CP$ violation in all charmonium modes.}
  \label{tab:charmoniumresult}
  \begin{tabular}{c|c|c}  
\hline
Decay Mode  &   $\mathcal{S}$  & $\mathcal{A}$ \\
\hline
$B^0 \to J/\psi K_S^0$ & $0.671\pm 0.029$ & $-0.014\pm0.021$\\
$B^0 \to J/\psi K_L^0$ & $-0.641\pm 0.047$ & $0.019 \pm 0.026$\\
$B^0 \to \psi' K_S^0$ & $0.739\pm0.079$ & $0.103\pm0.055$\\
$B^0 \to \chi_{c1} K_S^0$ & $0.636\pm0.117$ & $-0.023\pm0.083$\\
\hline
\end{tabular}
\end{center}
\end{table}

\begin{figure}[htbp]
\begin{center}
\includegraphics[width=3cm]{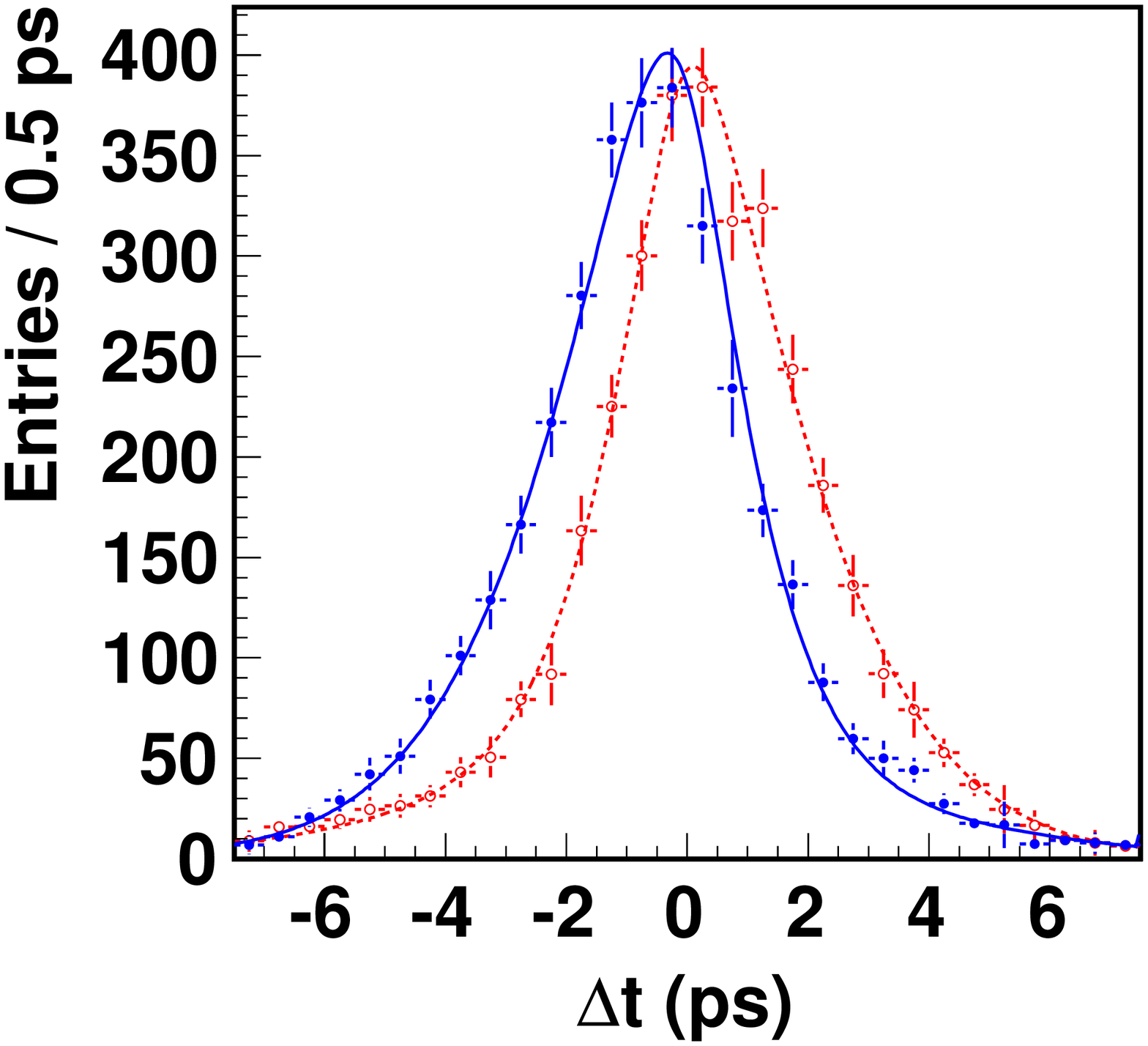}
\includegraphics[width=3cm]{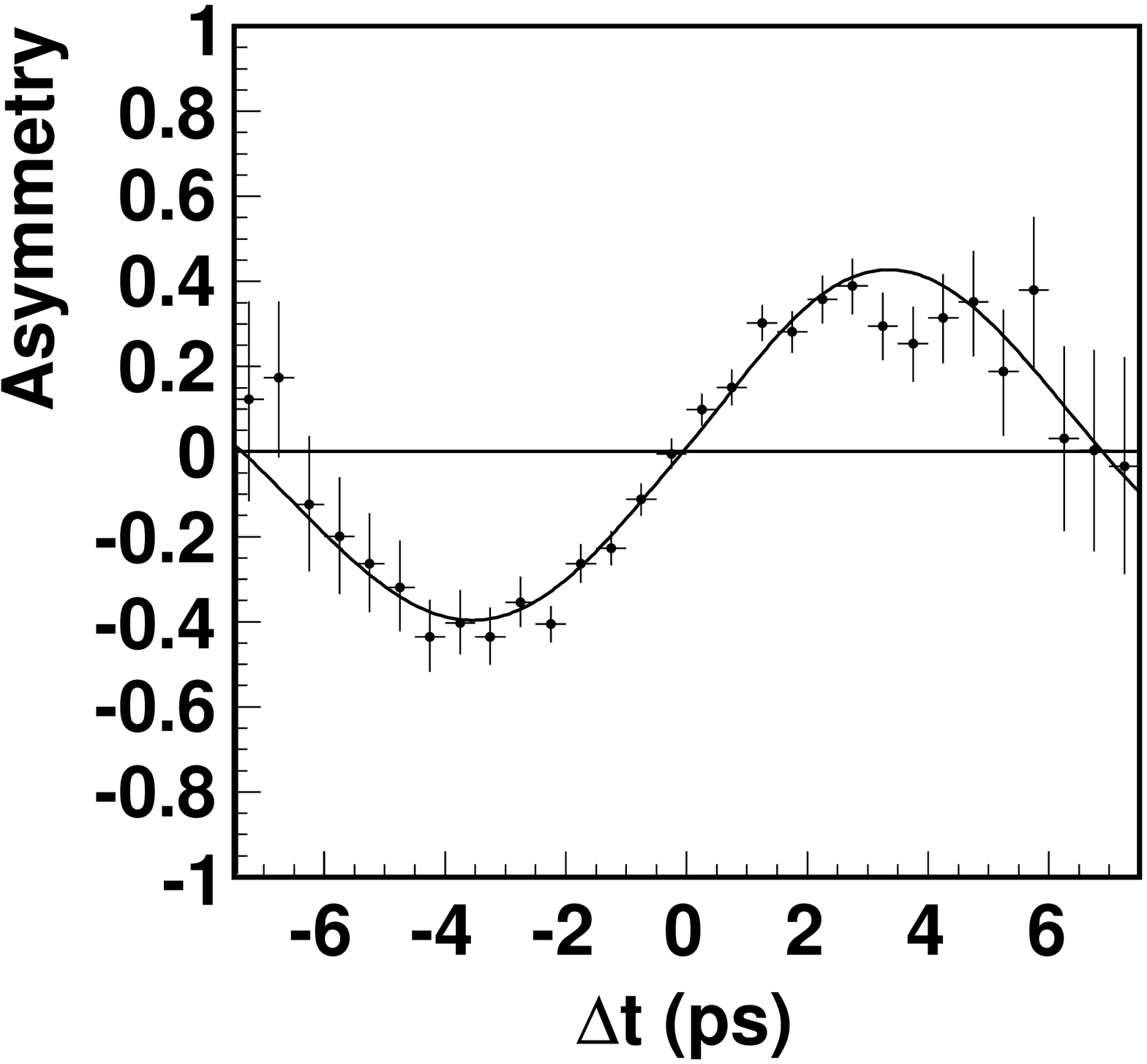}\\
\includegraphics[width=3cm]{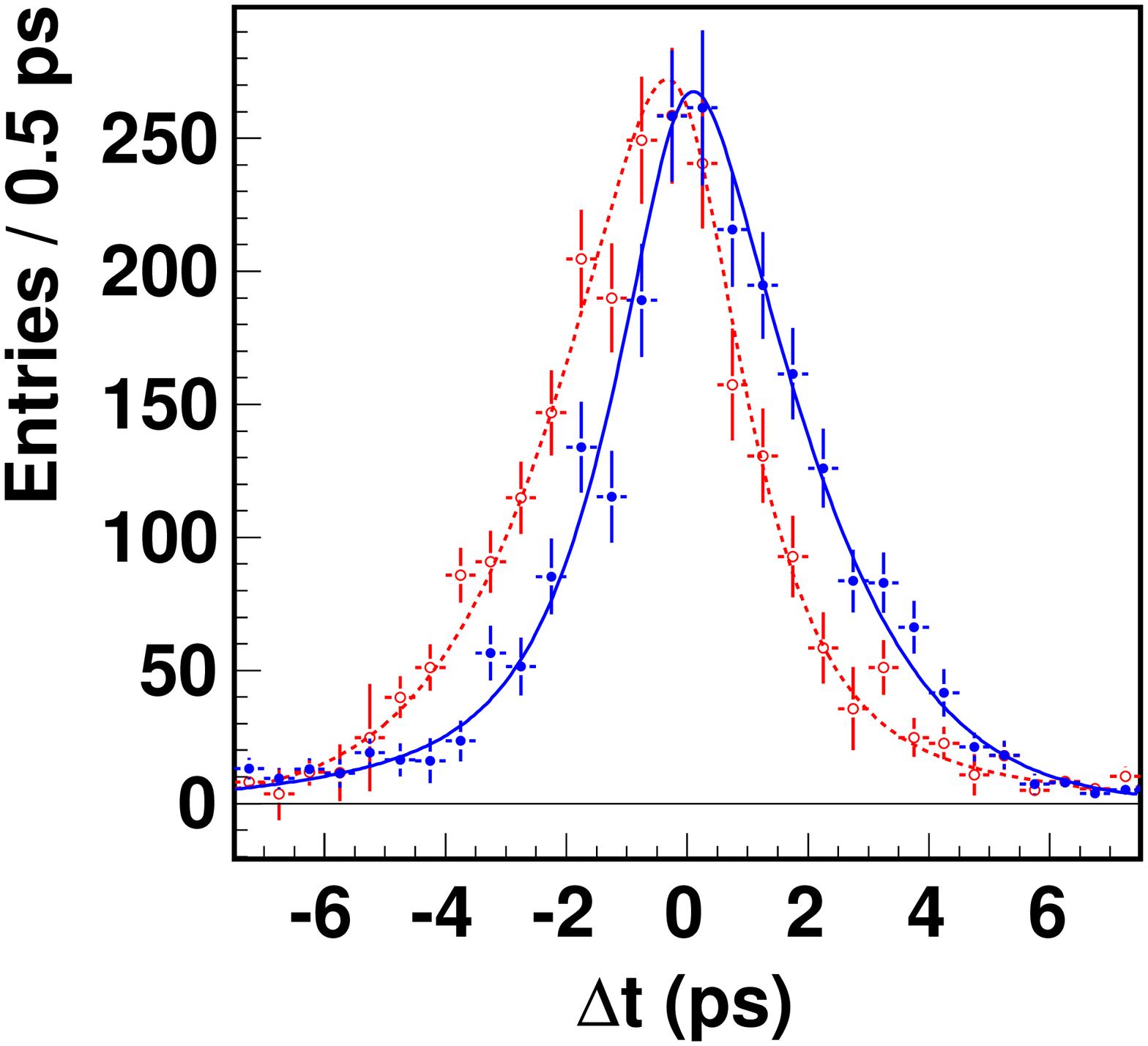}
\includegraphics[width=3cm]{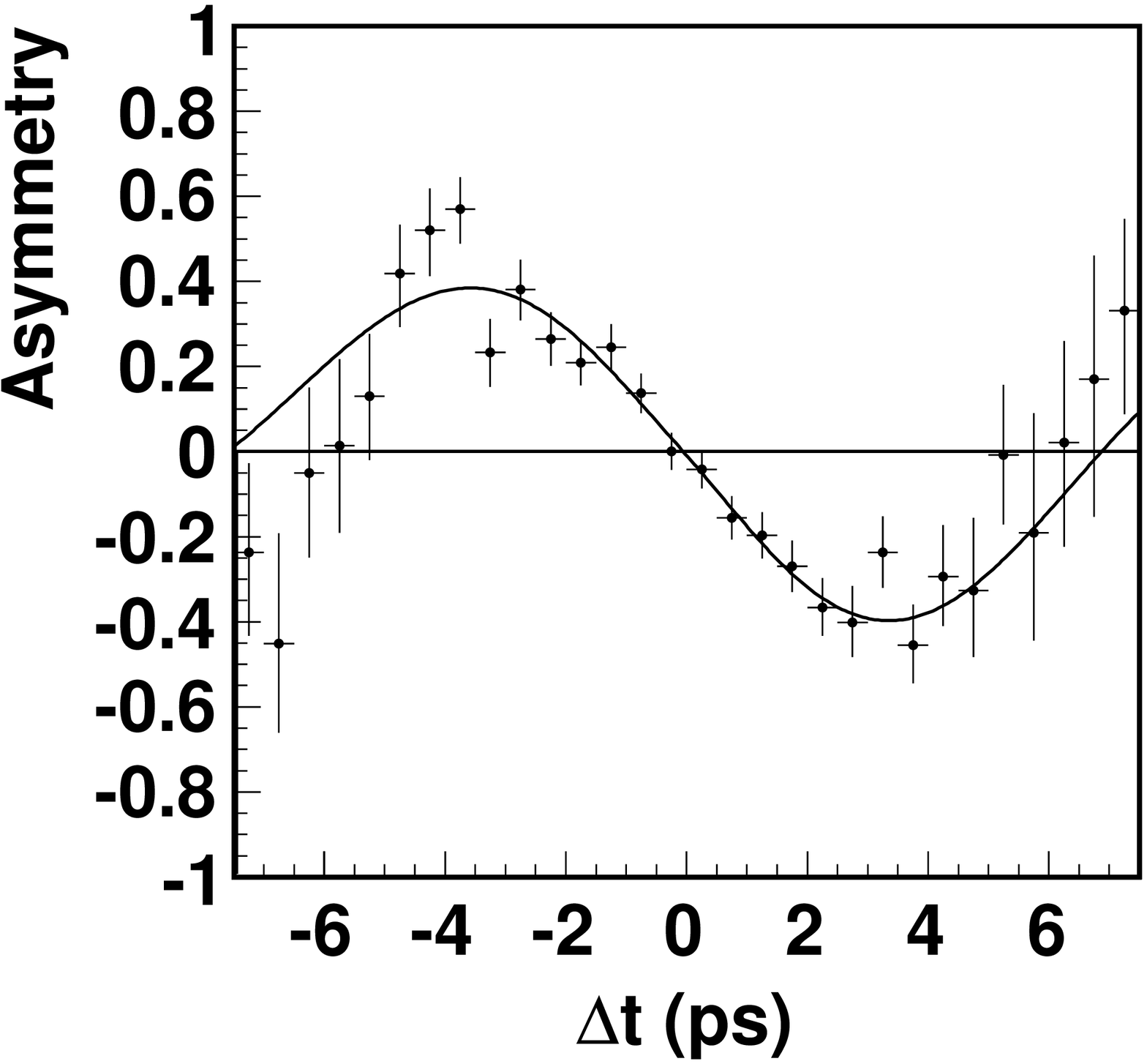}
\caption{$\Delta t$ distributions for $q=+1$ (red) and $q=-1$ (blue) and the raw asymmetry plots for $CP$-odd states (upper) and a $CP$-even ($B^0 \to J/\psi K_L^0$) state (lower). These are background-subtracted and for good-tagged events only.
The opposite $CP$ asymmetry of $B^0 \to J/\psi K_L^0$ is clearly visible from the lower plots.}
\label{fig:belledt}
\end{center}
\end{figure}

Figure~\ref{fig:belledt} shows the background-subtracted
$\Delta t$ distributions and the raw asymmetry 
for good-tagged events only ($r>0.5$).
We combine the $CP$-odd states,
$B^0 \to J/\psi K_S^0$, $B^0 \to \psi' K_S^0$ 
and $B^0 \to \chi_{c1} K_S^0$ together.
We define the raw asymmetry in each $\Delta t$ bin as 
$(N_{+}-N_{-})/(N_{+}+N_{-})$, where $N_{+}$ $(N_{-})$
is the number of observed candidates with $q=+1$ $(-1)$.
The systematic uncertainties are improved 
compared to previous Belle measurements~\cite{kfchen,sahoo}.

Combining all charmonium modes, Belle reported the
world's most precise measurements:
\begin{eqnarray}
\sin2\phi_1 = 0.668\pm0.023 \pm 0.013,
\nonumber\\
{\cal A} = 0.007\pm0.016 \pm 0.013,
\end{eqnarray}
where the uncertainties are statistical and systematic, respectively.

\begin{figure}[htbp]
\begin{center}
\includegraphics[width=6cm]{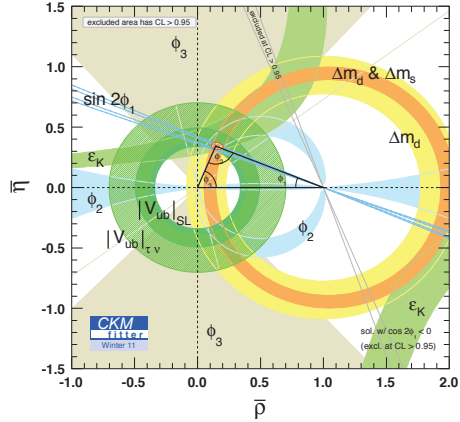}
\caption{The global fit result from the CKM fitter using all the recent measurements of the unitary triangle. }
\label{fig:ckmfitter}
\end{center}
\end{figure}

Combining this new  measurements 
from Belle with BaBar~\cite{babar-btoccs}, the 
new world average calculated
by the Heavy Flavor Averaging Group (HFAG)~\cite{hfag} is:
\begin{eqnarray}
\sin2\phi_1 (b \to c\bar{c}s) = 0.678\pm0.020,
\nonumber\\
{\cal A} (b \to c\bar{c}s) = -0.013\pm0.017.
\end{eqnarray}
The experimental uncertainty on $\sin 2\phi_1$ is reduced to $3\%$ and 
thus serves as a firm reference point for the SM. 
The value of ${\cal A}$ is consistent with zero.
The new results provide 
a better constraint on the allowed region in the CKM fitter as shown
in Fig.~\ref{fig:ckmfitter} and
give the value of $\phi_1(\beta)$ to be~\cite{hfag}
\begin{eqnarray}
\phi_1 (\beta) = (21.4\pm0.8)^{\circ},
\end{eqnarray}
which is the most precise measurement having an error $<1^{\circ}$.
Details on the measurements of the CKM angle $\phi_1/\beta$ 
at the $B$ factories are described
in the FPCP 2011 proceeding~\cite{sahoo-fpcp}.

\section{{\boldmath $b \to c\bar{c}d$} Decay Modes}

Belle updated the measurements of $CP$ violation parameters in
$b \to c \bar{c} d$ decays, 
$B^0 \to D^+D^-$ and $B^0 \to D^{*+}D^{*-}$,
using the full data sample.
The dominant contribution in these decays is from the tree-diagram.
If this is the only contribution, then 
SM expectation is ${\cal S} = -\sin 2\phi_1$ and ${\cal A} = 0$.
But, penguin contributions are expected to change the values 
by a few percent~\cite{xing-btoccbard}.
Large deviations from $\sin 2\phi_1$ in 
$b \to c\bar{c} s$ decays will be clear hint of NP.

\subsection{{\boldmath $B^0 \to D^+D^-$}}

\begin{figure}[htbp]
\begin{center}
\includegraphics[width=4cm]{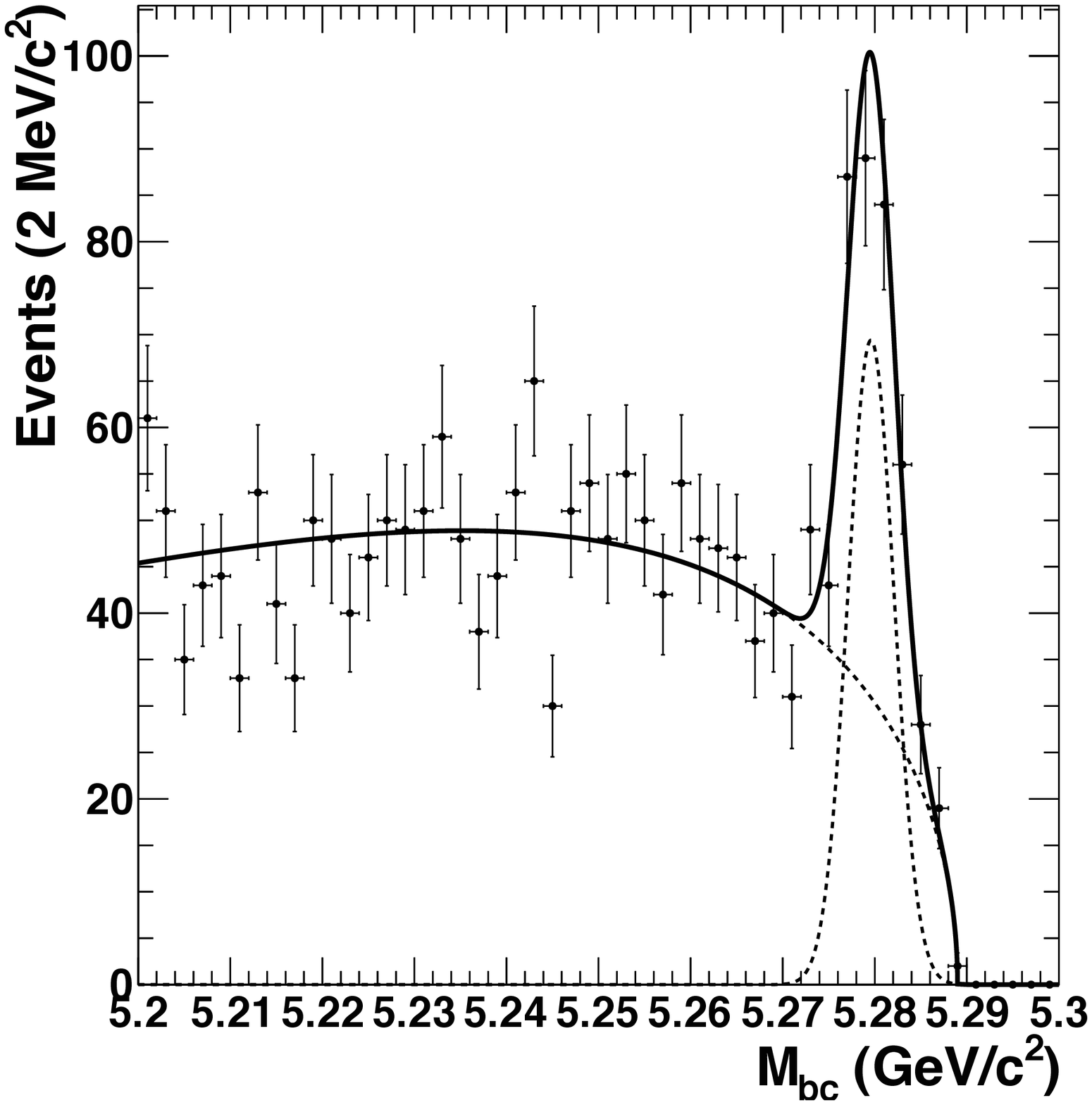}
\includegraphics[width=4cm]{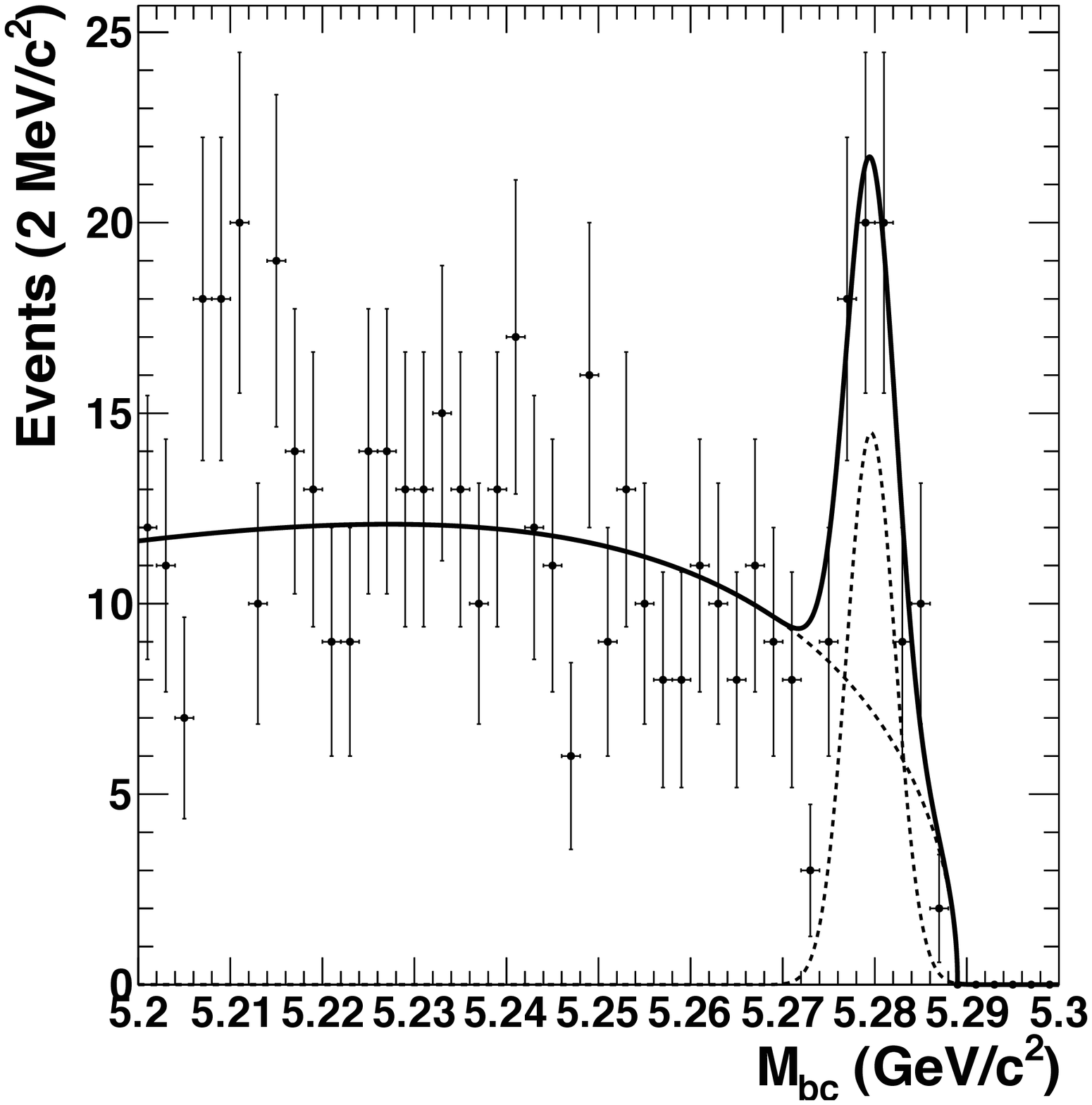}
\includegraphics[width=4cm]{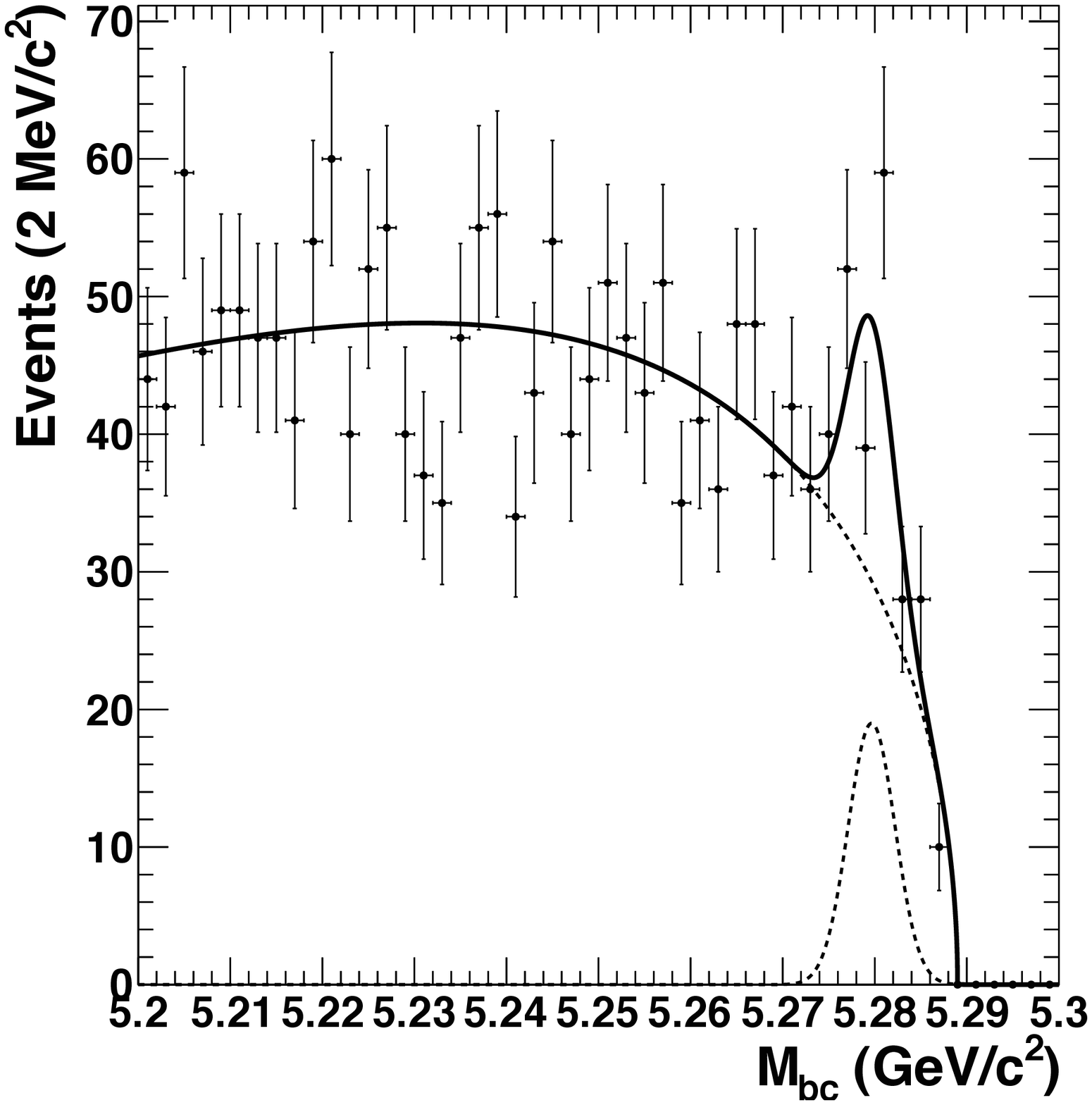}
\caption{The $B^0 \to D^+D^-$ $M_{\rm bc}$ distributions for $K\pi\pi$ (left), $K_S^0\pi$ (middle) and $K_S^0\pi\pi^0$ mode (right).}
\label{fig:mbc-dplusdminus}
\end{center}
\end{figure}

The $D$ mesons are reconstructed using the 
$D^+ \to K^-\pi^+\pi^+$, $D^+ \to K_S^0 \pi^+$ 
and $D^+ \to K_S^0 \pi^+ \pi^0$ decays~\cite{conjugate}.
Signal is extracted from an 
extended UML fit to the
two-dimensional $\Delta E$-$M_{\rm bc}$ distribution.
The projections of the fit onto $M_{\rm bc}$ for each mode are
shown in Fig.~\ref{fig:mbc-dplusdminus}.
Contributions from $B^0 \to D^+K^{(*)0}\pi^-$ peaking
background are estimated using a $D$ mass sideband in data
and are subtracted from the signal yield.
The signal yields and measured branching fractions for each channel
are summarized in the Table~\ref{tab:BrDplusDminus}.
\begin{table}[hbtp]
\begin{center}
\caption{Signal yields and measured branching fractions in each channel for the $B^0 \to D^+D^-$ mode.}
\label{tab:BrDplusDminus}
\begin{tabular}{c|c|c}  
\hline
Decay Mode  &  Signal Yield  & Branching Fraction \\
\hline
$(K^-\pi^+\pi^+)(K^+\pi^-\pi^-)$ &  $221.4\pm18.6$  & $(2.16\pm0.18) \times 10^{-4}$ \\
$(K^-\pi^+\pi^+)(K_S^0\pi^-)$ &  $48.0\pm8.9$  & $(1.96\pm0.36) \times 10^{-4}$ \\
$(K^-\pi^+\pi^+)(K_S^0\pi^-\pi^0)$ &  $54.1\pm14.6$  & $(1.83\pm0.49) \times 10^{-4}$ \\
\hline
\end{tabular}
\end{center}
\end{table}

Combining the three decay modes, we measure the branching fraction to be
\begin{eqnarray}
{\mathcal B}(B^0 \to D^+D^-) = (2.09\pm 0.15 \pm 0.18) \times 10^{-4}.
\end{eqnarray}
This measurement is consistent with Belle's previous result using $535 \times 10^6$ $B\bar{B}$ pairs~\cite{sasa-DplusDminus}.
The dominant contributions to the systematic uncertainty 
come from the $K/\pi$ selection efficiency,
daughter branching fractions and continuum suppression.

We use the $K\pi\pi$ and $K_S^0\pi$ modes in the 
time-dependent measurement.
These modes show no significant peaking background in contrast to the $K_S^0\pi\pi^0$ mode.
The measured $CP$ violation parameters are
\begin{eqnarray}
{\mathcal S} = -1.06 \pm 0.21 (\rm stat) \pm 0.07 (\rm syst) \; {\rm and} \nonumber \\
{\mathcal A} = +0.43 \pm 0.17 (\rm stat) \pm 0.04 (\rm syst).
\end{eqnarray}

\begin{figure}[htbp]
\begin{center}
\includegraphics[width=8cm]{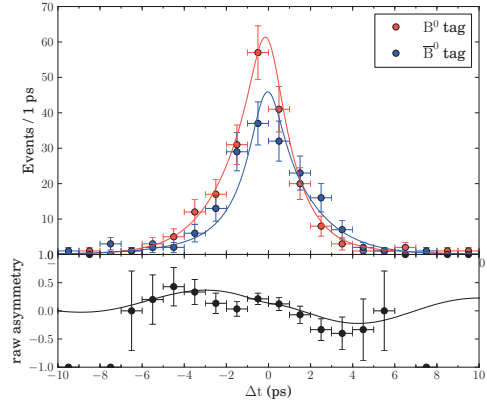}
\caption{$\Delta t$ distributions for $q=+1$ (red) and $q=-1$ (blue) (upper) and the raw asymmetry (lower) of the combined $CP$ fit of
$k\pi\pi$ and $K_S^0\pi$ modes in $B^0 \to D^+D^-$.}
\label{fig:deltat-dplusdminus}
\end{center}
\end{figure}

The ${\cal S}$ value is consistent with the 
measurement from $b \to c\bar{c}s$ decays.
Compared to the previous Belle publication~\cite{sasa-DplusDminus} 
the direct $CP$ violation parameter (${\cal A}$)
decreased and its central value moved towards the expected value, 
which is close to zero.
The $\Delta t$ distributions and raw asymmetry for $B^0 \to D^+D^-$ 
are shown in the Fig.~\ref{fig:deltat-dplusdminus}.

\subsection{{\boldmath $B^0 \to D^{*+}D^{*-}$}}

Belle also updated the measurements of the 
branching fraction, polarization and $CP$ violation parameters
for $B^0 \to D^{*+}D^{*-}$ using the full data sample.
The signal is reconstructed in a total of nine hadronic $D$ decay modes.
The signal yield is obtained from a two-dimensional extended UML fit to the $\Delta E$-$M_{\rm bc}$
distribution.
We obtain a signal yield of $1225\pm59$ events and the measured branching fraction is
\begin{eqnarray}
{\mathcal B}(B^0 \to D^{*+}D^{*-}) = (7.82\pm 0.38 \pm 0.60) \times 10^{-4}.
\end{eqnarray}
We have a large gain in signal yield compared to 
Belle's previous publication with 
$657 \times 10^6$ $B\bar{B}$ pairs~\cite{kim-DstpDstm} 
due to higher track multiplicity in the final state.

\begin{figure}[htbp]
\begin{center}
\includegraphics[width=6cm]{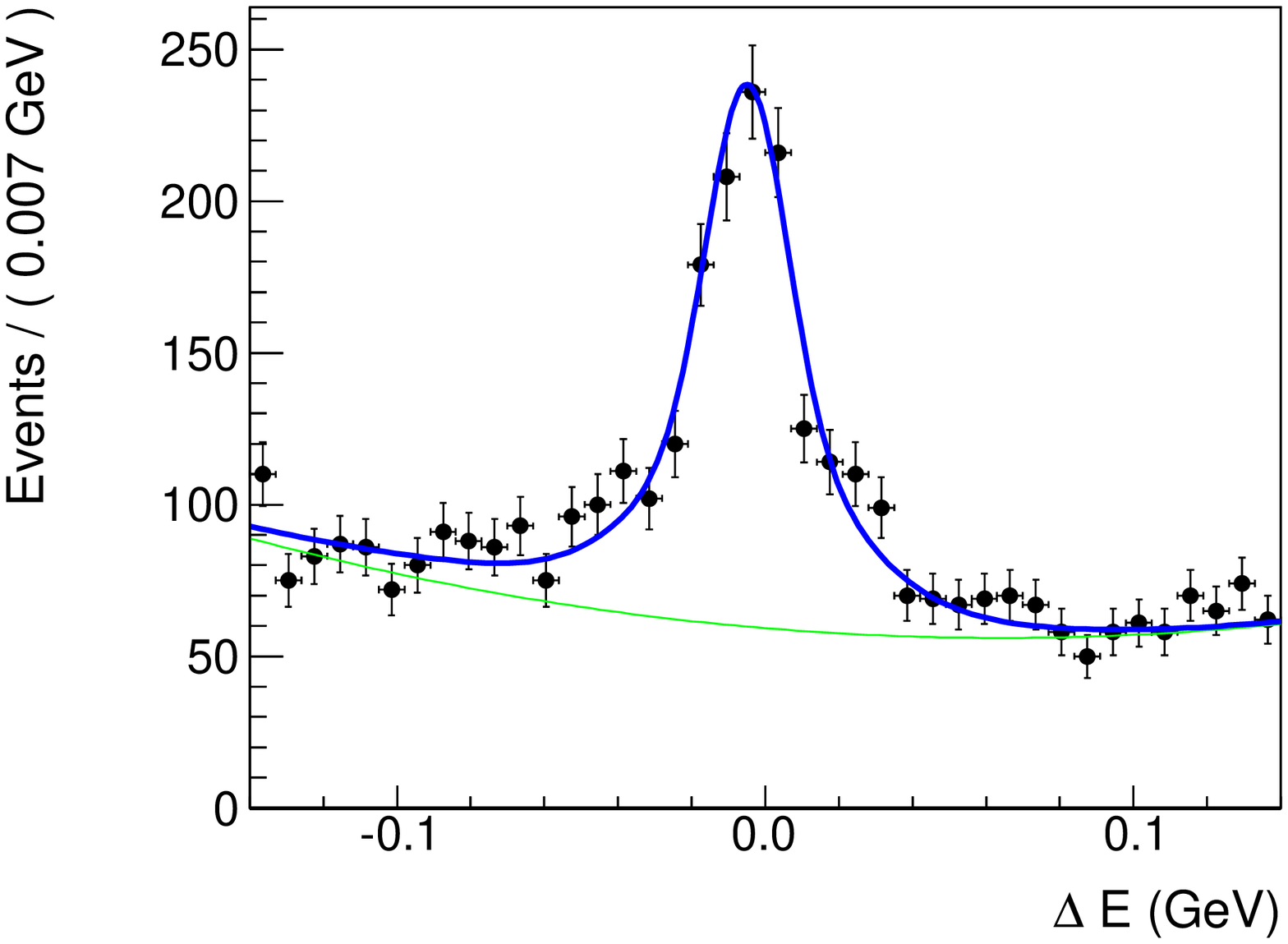}
\includegraphics[width=6cm]{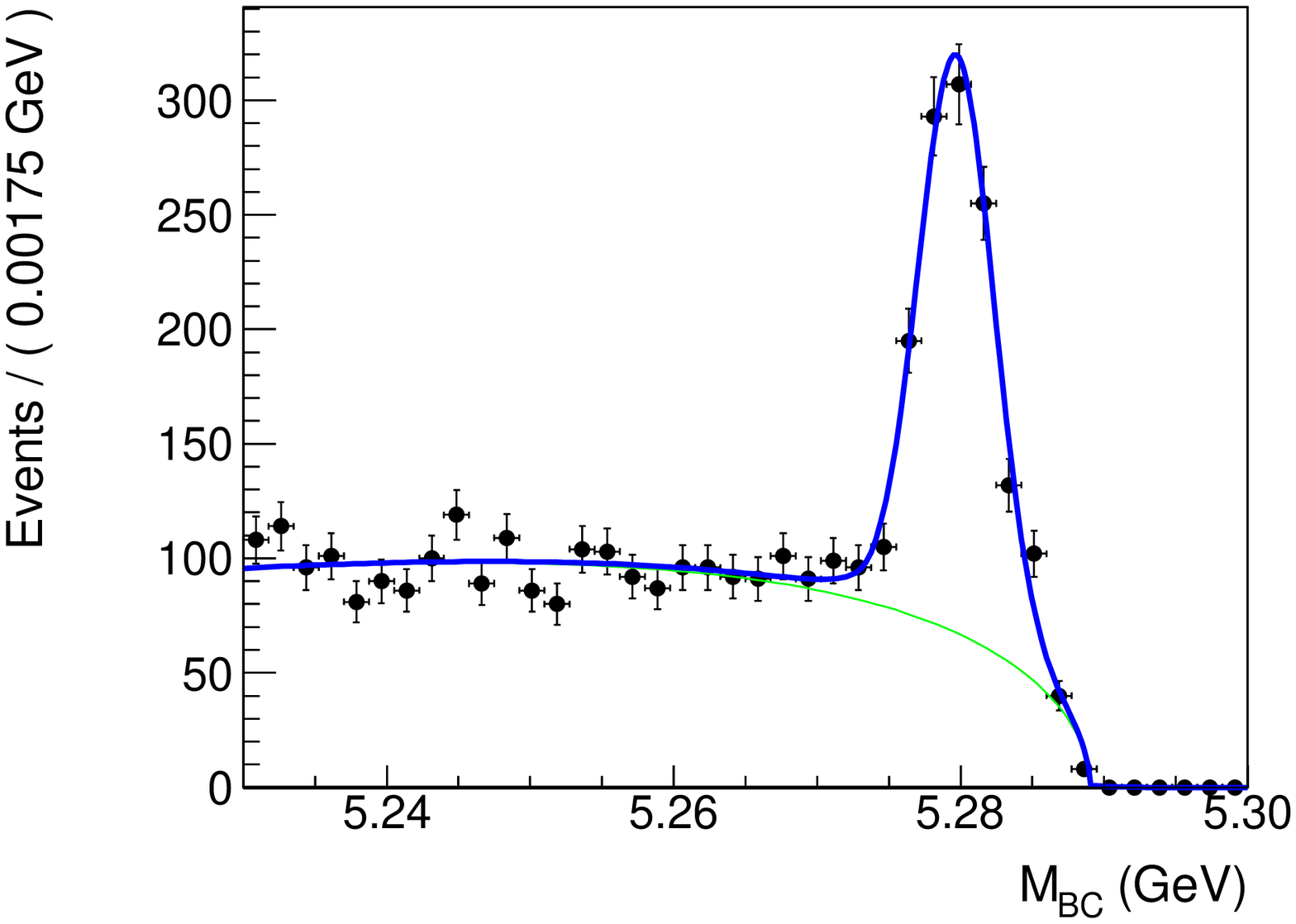}
\caption{$\Delta E$ (left) and $M_{\rm bc}$ (right) distributions for $B^0 \to D^{*+}D^{*-}$ mode.}
\label{fig:dembc-ddstar}
\end{center}
\end{figure}

\begin{figure}[htbp]
\begin{center}
\includegraphics[width=6cm]{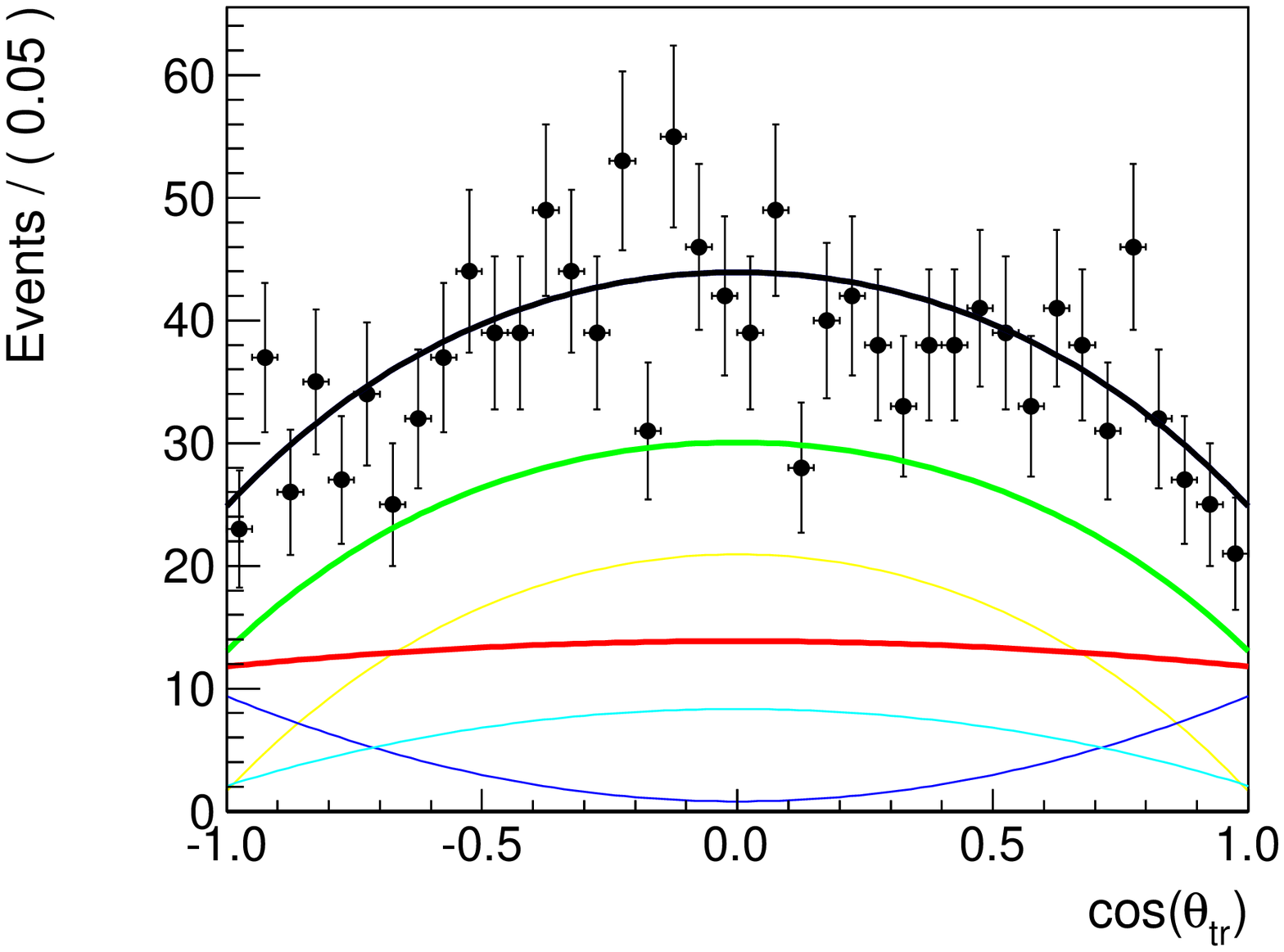}
\includegraphics[width=6cm]{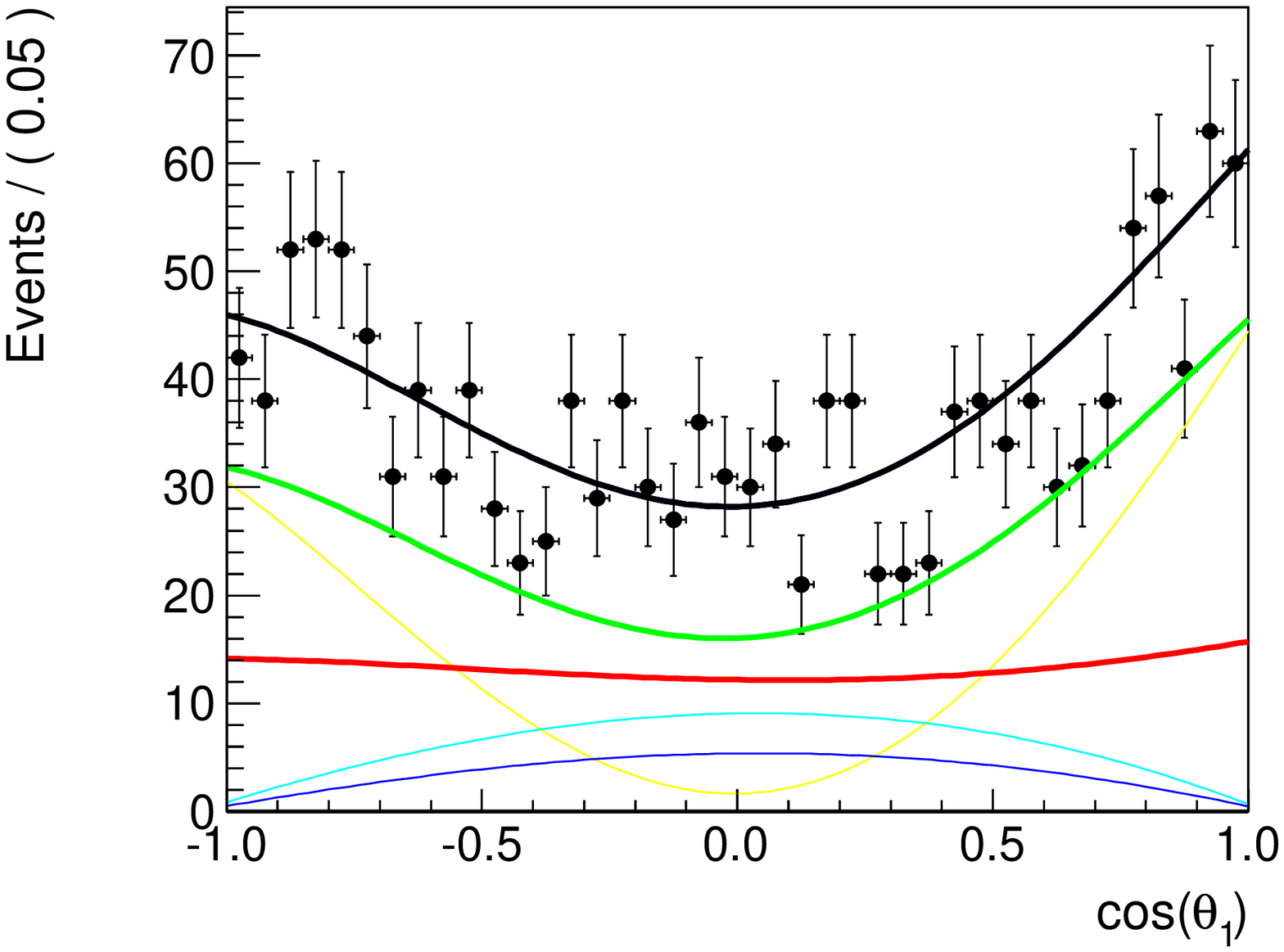}
\caption{The projections of the angular distributions in the signal region of $\Delta E$ and $M_{\rm bc}$. The black (green, red) line shows the total (signal, background) pdf, while orange, blue and cyan lines show the signal contributions for each of the polarization states.}
\label{fig:pol-ddstar}
\end{center}
\end{figure}

$B^0 \to D^{*+} D^{*-}$ is the decay of a scalar to two vector mesons; the final state is 
a mixture of $CP$-even and $CP$-odd states.
Therefore, an angular analysis has to be performed to determine 
the fraction of $CP$-even decays.
This is done using two of the three possible angles of the transversity base, 
denoted by $\cos \theta_{tr}$ and $\cos \theta_1$.
The measurements of polarization and $CP$ violation are performed simultaneously with a
five dimensional fit to $\Delta E$, $M_{\rm bc}$, $\cos \theta_{tr}$, 
$\cos \theta_1$ and $\Delta t$ distributions.
The $\cos \theta_{tr}$ and $\cos \theta_1$ distributions for data in the 
signal region of $\Delta E$-$M_{\rm bc}$ are shown in Fig.~\ref{fig:pol-ddstar}.

We measure the polarization and $CP$ violation parameters to be
\begin{eqnarray}
{\mathcal S}         = -0.79 \pm 0.13 \pm 0.03 \nonumber \\
{\mathcal A}         = +0.15 \pm 0.08 \pm 0.02 \nonumber \\
{\mathcal R}_0       = 0.62 \pm 0.03 \pm 0.01 \nonumber \\
{\mathcal R}_{\perp} = 0.14 \pm 0.02 \pm 0.01 
\end{eqnarray}

Both the measurements of branching fraction and the polarization are 
consistent with Belle's previous measurement~\cite{kim-DstpDstm}.
The $CP$ violation parameters are consistent with SM predictions.

\begin{figure}[htbp]
\begin{center}
\includegraphics[width=6cm]{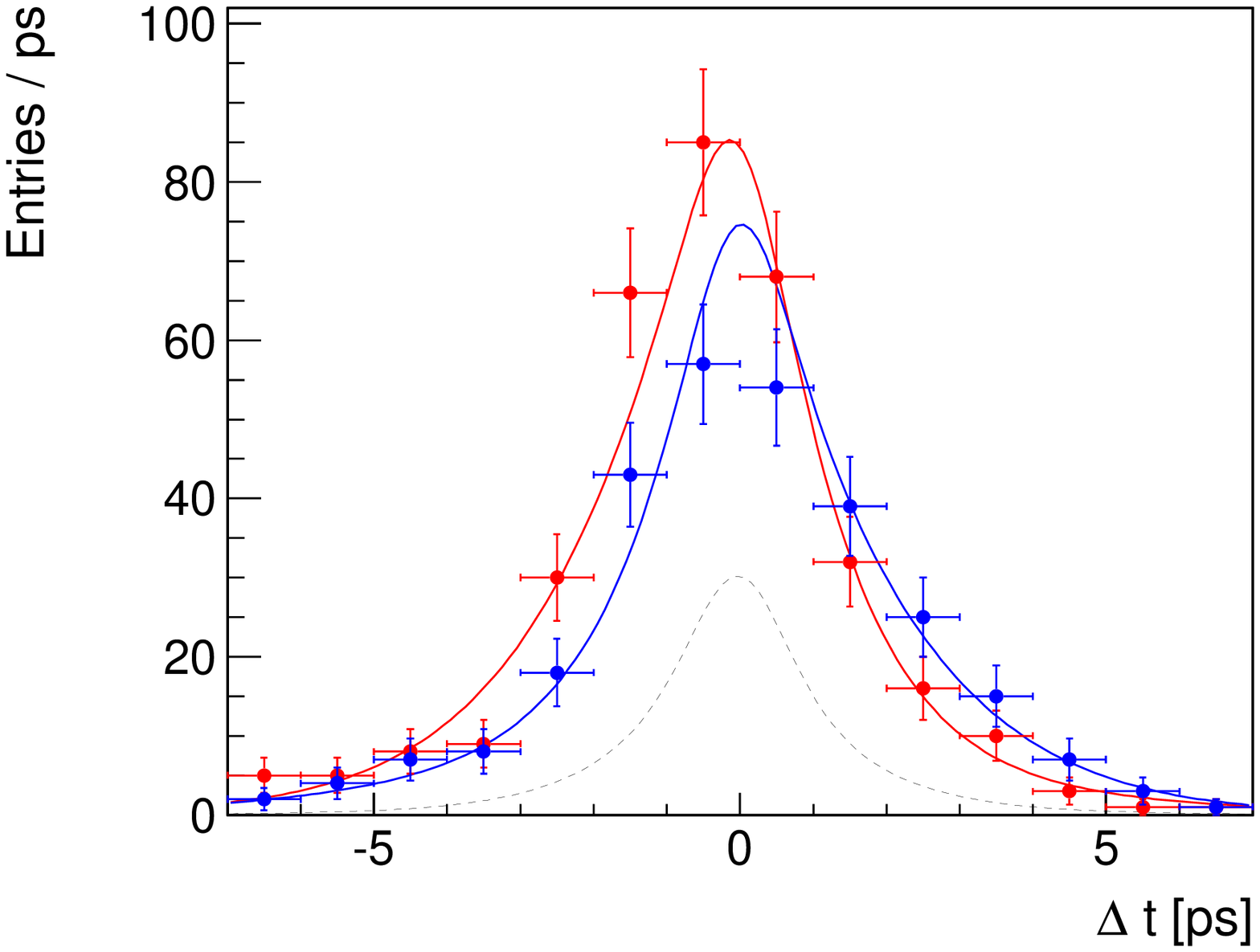}
\includegraphics[width=8cm]{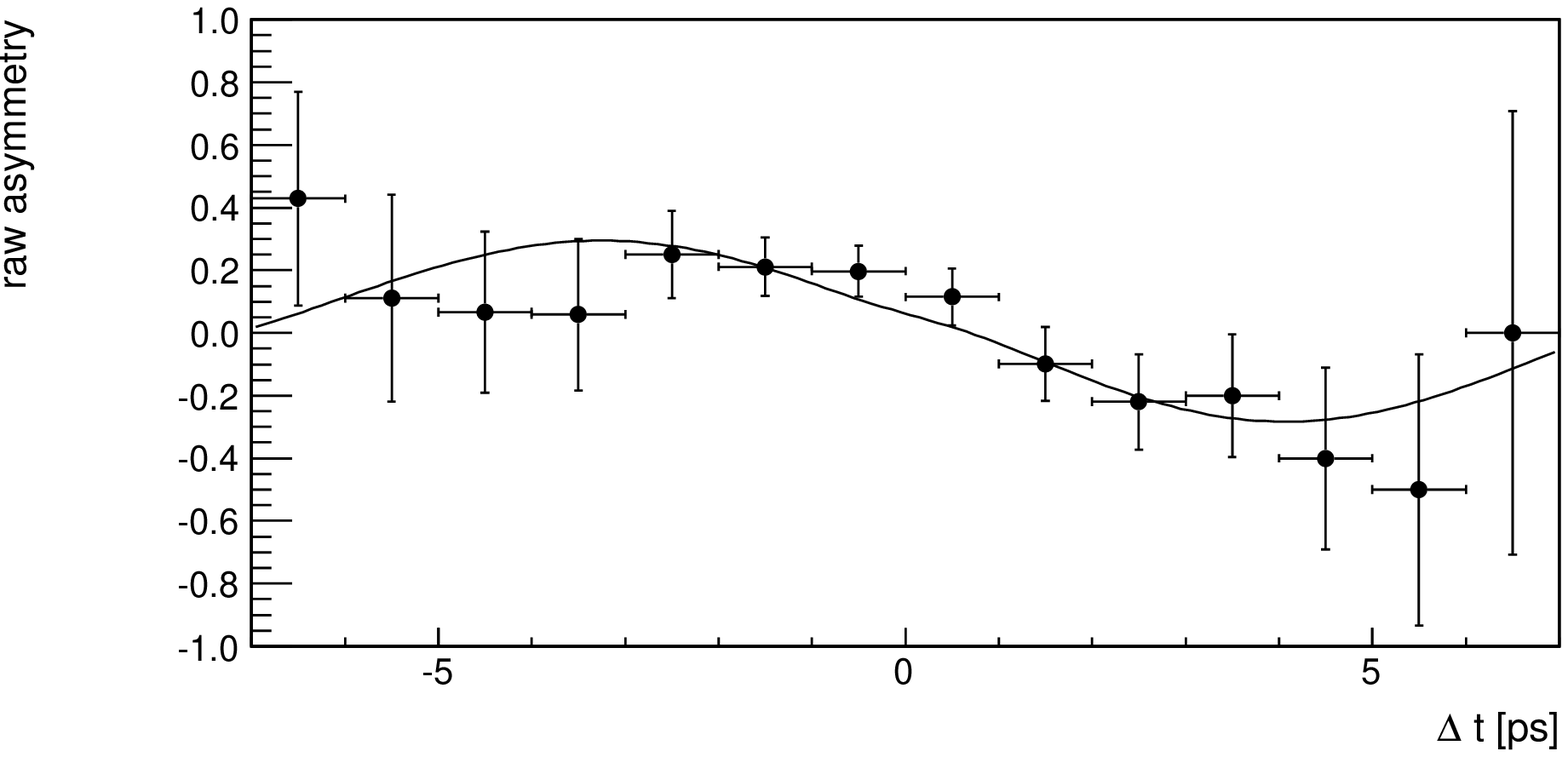}
\caption{$\Delta t$ distributions for $q=+1$ (red) and $q=-1$ (blue) (left) and the raw asymmetry plot (right) for $B^0 \to D^{*+} D^{*-}$ mode. These are for good-tagged events only.}
\label{fig:deltat-dstdst}
\end{center}
\end{figure}

\begin{figure}[htbp]
\begin{center}
\includegraphics[width=5cm]{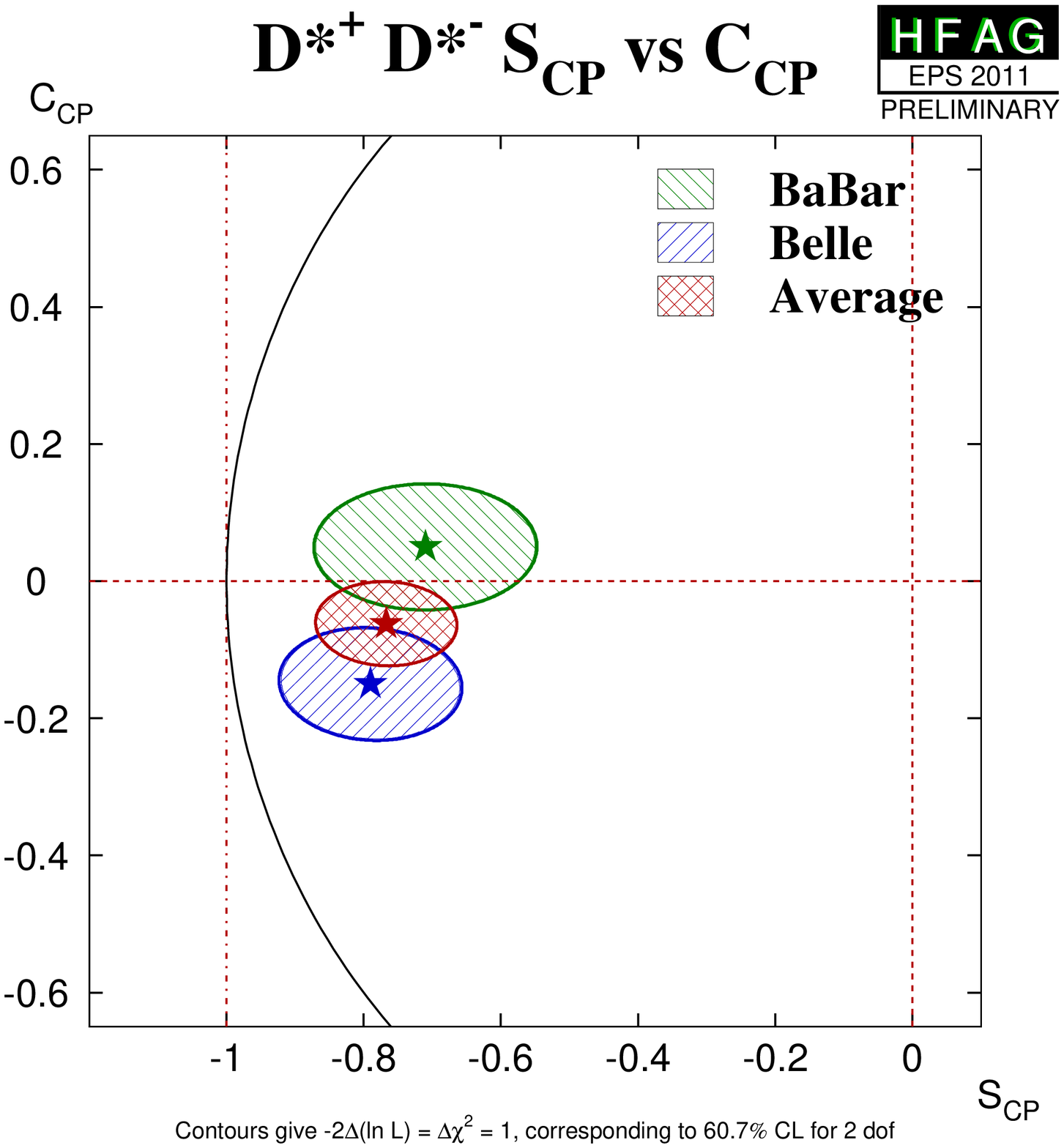}
\includegraphics[width=5cm]{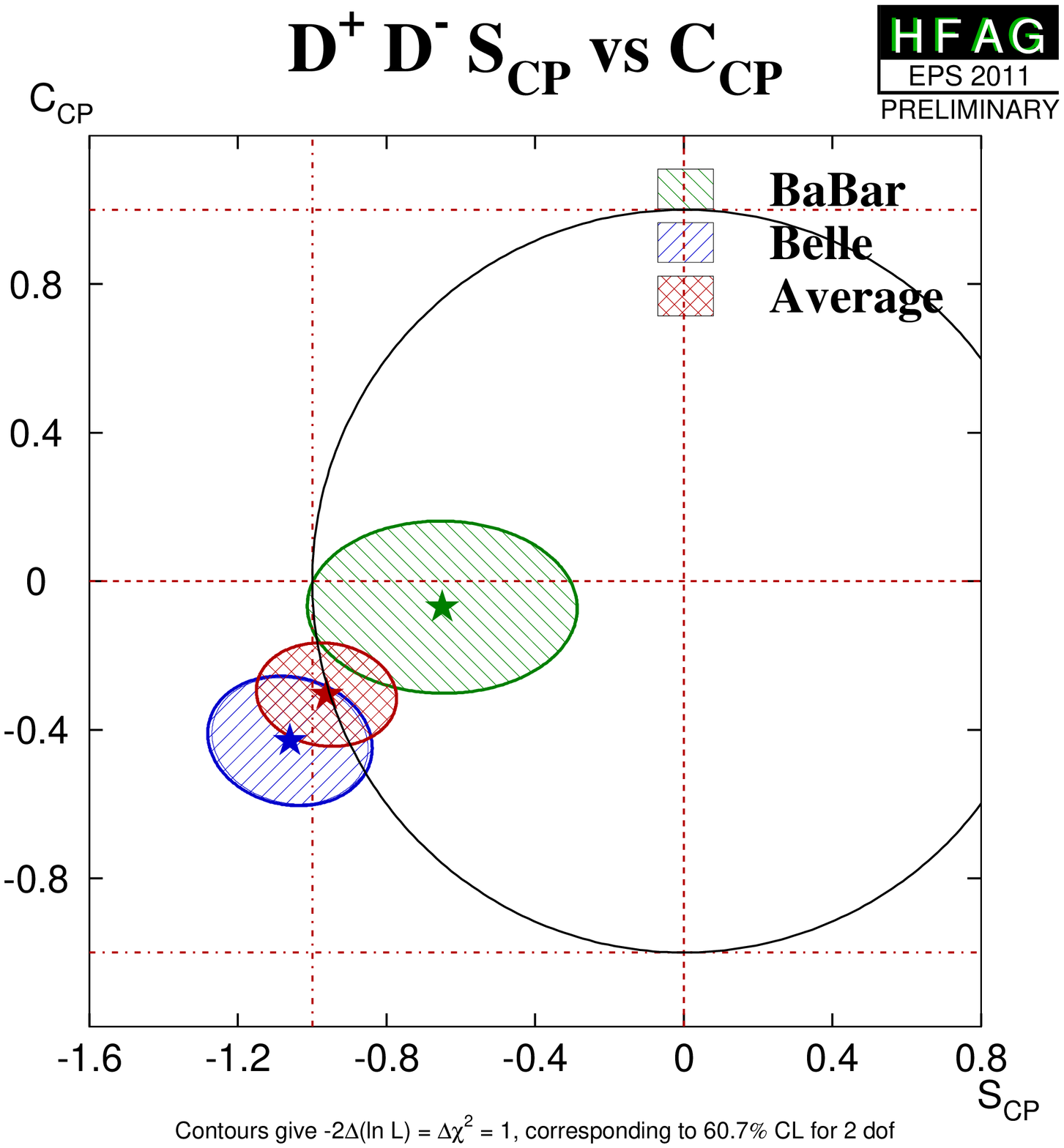}
\caption{The HFAG summary of the results on $B^0 \to D^+D^-$ and $B^0 \to D^{*+}D^{*-}$ from $B$ factories.}
\label{fig:hfag-btoccd}
\end{center}
\end{figure}

The HFAG summary of the $b \to c\bar{c}d$ results from Belle and BaBar
are shown in Fig.~\ref{fig:hfag-btoccd}.

\section{ {\boldmath CPT} violation parameters in {\boldmath $B^0$} decay}

$CPT$ conservation is one of the most fundamental laws in the SM.
Any evidence of $CPT$ violation will be clear hint of NP.
Using a data sample of $535 \times 10^6$ $B\overline{B}$ pairs, 
we measure the
$CPT$ violating parameter in 
$B^0 \to J/\psi K_S^0$, $B^0 \to J/\psi K_L^0$,
$B^0 \to D^- \pi^+$, $B^0 \to D^{*-} \pi^+$, $B^0 \to D^{*-} \rho^+$ and
$B^0 \to D^{*-} l^+ \nu_l$ decays.
Assuming $CPT$ violation, we can express the mass eigenstates of the $B$ meson 
as a linear combination of weak eigenstates by
introducing a complex number $z$ in $B^0$-$\overline{B}{}^0$ mixing.
Any non-zero value of $z$ (in either the real or the imaginary term) 
will indicate $CPT$ violation in mixing.
The decay modes $B^0 \to J/\psi K_S^0$ and $B^0 \to J/\psi K_L^0$ 
are sensitive to ${\cal R}e(z)$ and $\Delta \Gamma_d/\Gamma_d$, while
the modes
$B^0 \to D^- \pi^+$, $B^0 \to D^{*-} \pi^+$, $B^0 \to D^{*-} \rho^+$ and $B^0 \to D^{*-} l^+ \nu_l$ 
are sensitive to ${\cal I}m(z)$.
The charged modes $B^+ \to J/\psi K^+$ and $B^+ \to D^0 \pi^+$ are used to 
determine the $\Delta t$ resolution.
We have a total of 534,068 neutral $B$ and 248,775 charged $B$ events.

\begin{figure}[ht]
  \begin{center}
    \includegraphics[width=5cm]{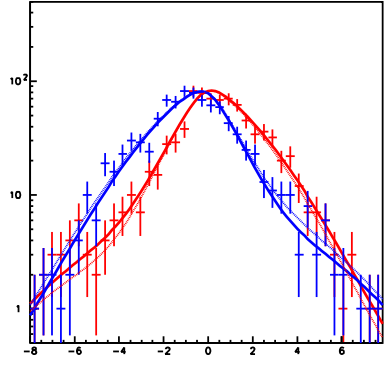}
    \caption{$\Delta t$ distributions for $q$ = $+1$ (red) and $q$ = $-1$ (blue) for $B^0 \to J/\psi K_S^0$ mode with good flavor tag events only. The light lines indicate ${\cal R}e(z) = 0.2$ and ${\cal I}m(z) = 0.0$.}
\label{fig:deltat-cpt}
\end{center}
\end{figure}

Using an UML fit to 72 free parameters, we measure the $CPT$ violating parameters as
\begin{eqnarray}
{\cal R}e(z) = (+1.9 \pm 3.7 \pm 3.2) \times 10^{-2}, \\ \nonumber
{\cal I}m(z) = (-5.7 \pm 3.3 \pm 6.0) \times 10^{-2}, \\ \nonumber
\Delta \Gamma_d/\Gamma_d = (-1.7 \pm 1.8 \pm 1.1) \times 10^{-2}.
\end{eqnarray}
The results are consistent with zero.
The $\Delta t$ distribution for $B^0 \to J/\psi K_S^0$ is shown in Fig.~\ref{fig:deltat-cpt}.

\section{First Observation of {\boldmath $B^0 \to \phi K_S^0 \gamma$} decay}

\begin{figure}[htbp]
  \begin{center}
    \includegraphics[width=6cm]{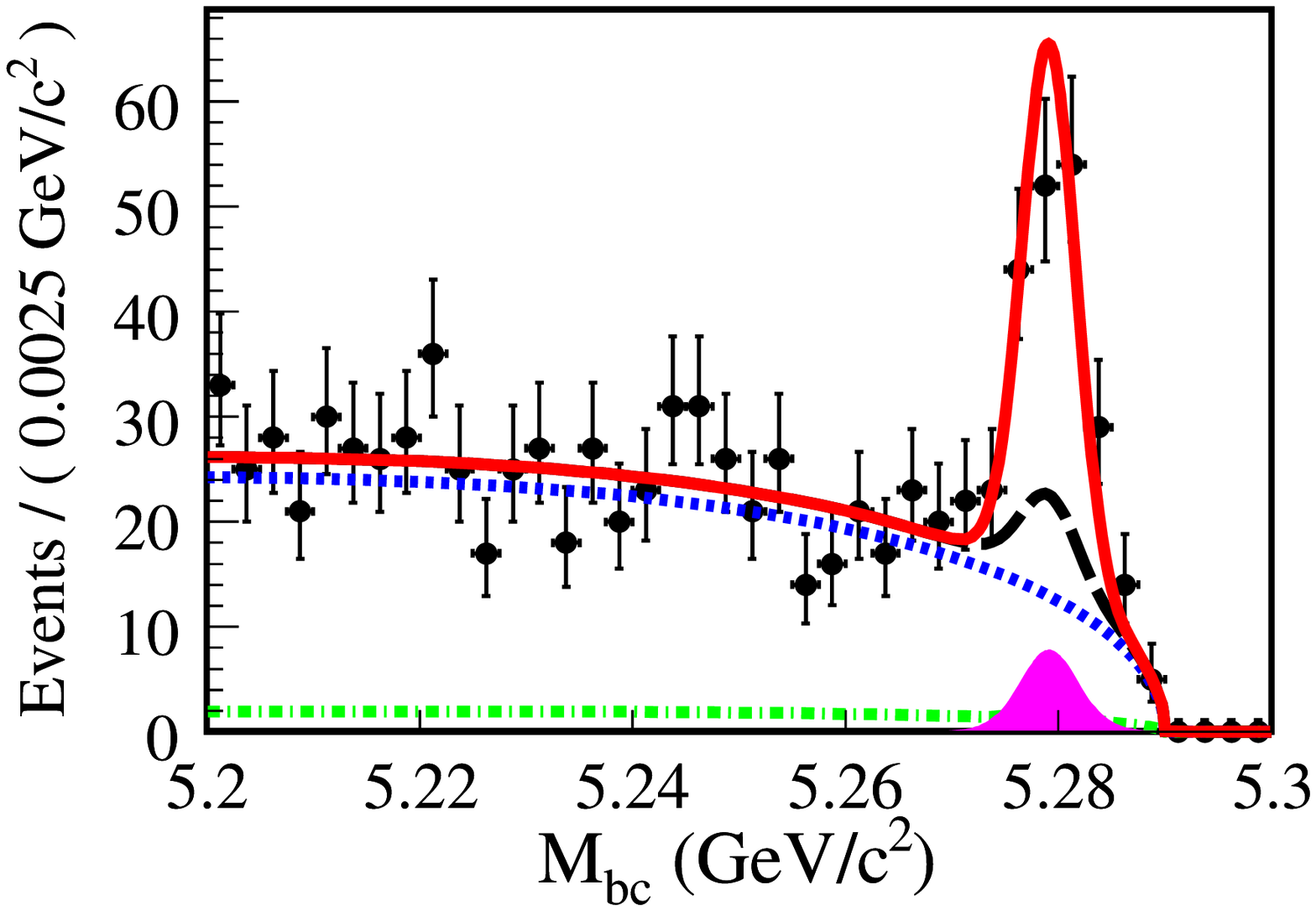}
    \includegraphics[width=6cm]{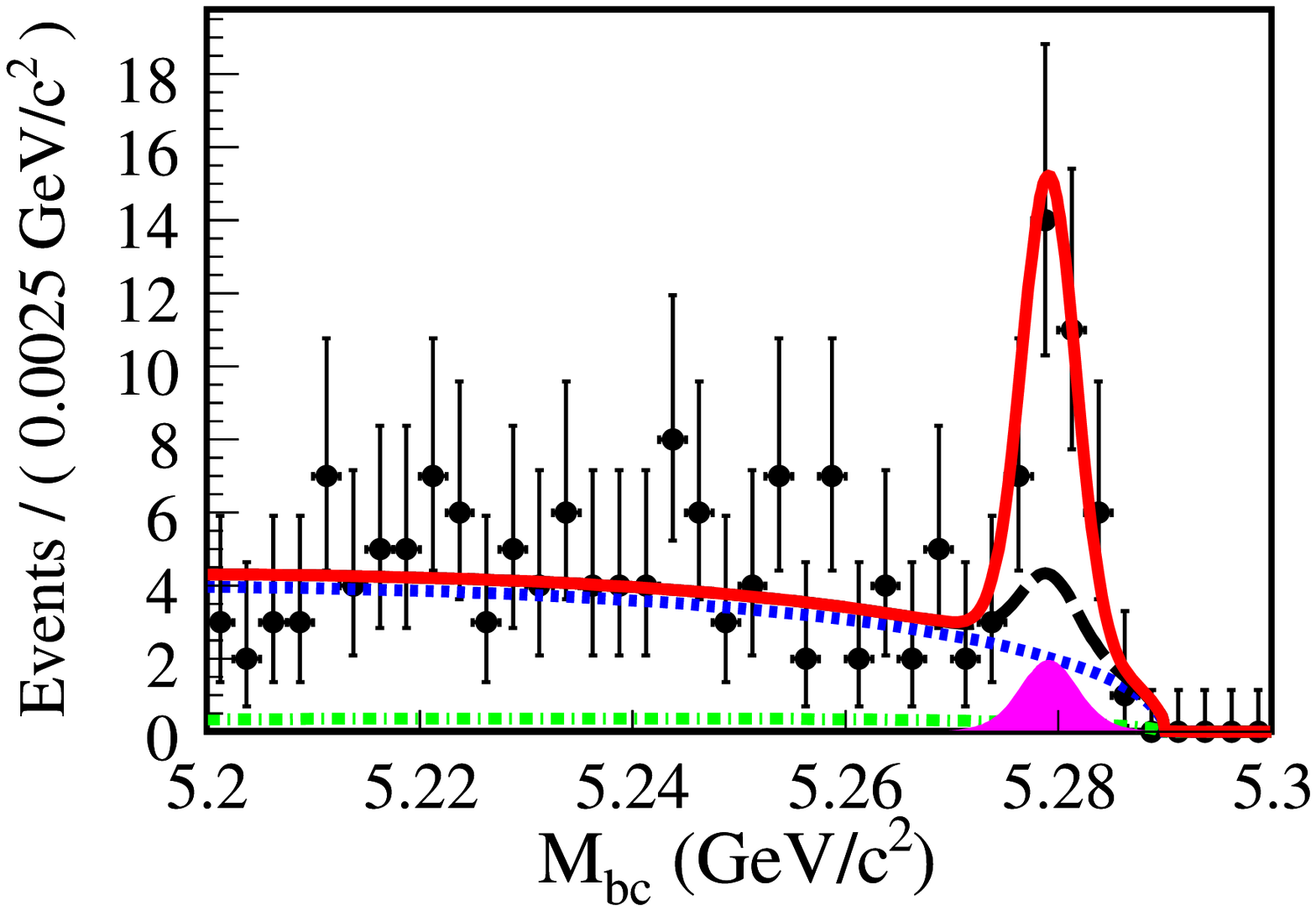}
    \caption{$M_{\rm bc}$ projections 
      for $B^+ \to \phi K^+ \gamma$ (left) and $B^0 \to \phi K_S^0 \gamma$ (right).
      The points with error bars are the data. The curves show the total fit function (solid red), 
      total background function (long-dashed black), continuum component (dotted blue), 
      the $b\rightarrow c$ component (dashed-dotted green) and the non-resonant 
      component as well as other charmless backgrounds (filled magenta histogram).
    }
    \label{fig:mbc-phikgamma}
  \end{center}
\end{figure}

Using the full data sample we observed for the first time the neutral mode
$B^0 \to \phi K_S^0 \gamma$~\cite{sahoo-phikgamma}.
We measure the branching fractions for both 
charged $B^+ \to \phi K^+ \gamma$ 
and neutral $B^0 \to \phi K^0 \gamma$, with 
$\phi \to K^+K^-$ and $K_S^0 \to \pi^+\pi^-$.
The signal yield is obtained from an extended
UML fit to the two-dimensional 
$\Delta E$-$M_{\rm bc}$ distribution.
The dominant background comes from 
$e^+e^- \rightarrow q\overline{q}$
($q = u, d, s, c$) 
continuum events. 
Another significant background is
non-resonant (NR) $B \rightarrow K^+ K^- K \gamma$, which peaks in the
$\Delta E$-$M_{\rm bc}$ signal region; it is estimated using
the $\phi$ mass sideband,
$M_{K^+K^-} \in \left[1.05,1.30\right]$ GeV/$c^2$, in data. 
The projections of the 
fit results onto $M_{\rm bc}$ are shown in 
Fig.~\ref{fig:mbc-phikgamma}. 
The fit yields a signal of 
$144\pm17$ $B^+ \to \phi K^+ \gamma$ 
and $37\pm8$ $B^0 \to \phi K_S^0 \gamma$ candidates.
The signals in the charged and neutral modes have significances of
$9.6\,\sigma$ and $5.4\,\sigma$, respectively.

\begin{figure}[htbp]
  \begin{center}
   \includegraphics[width=6cm]{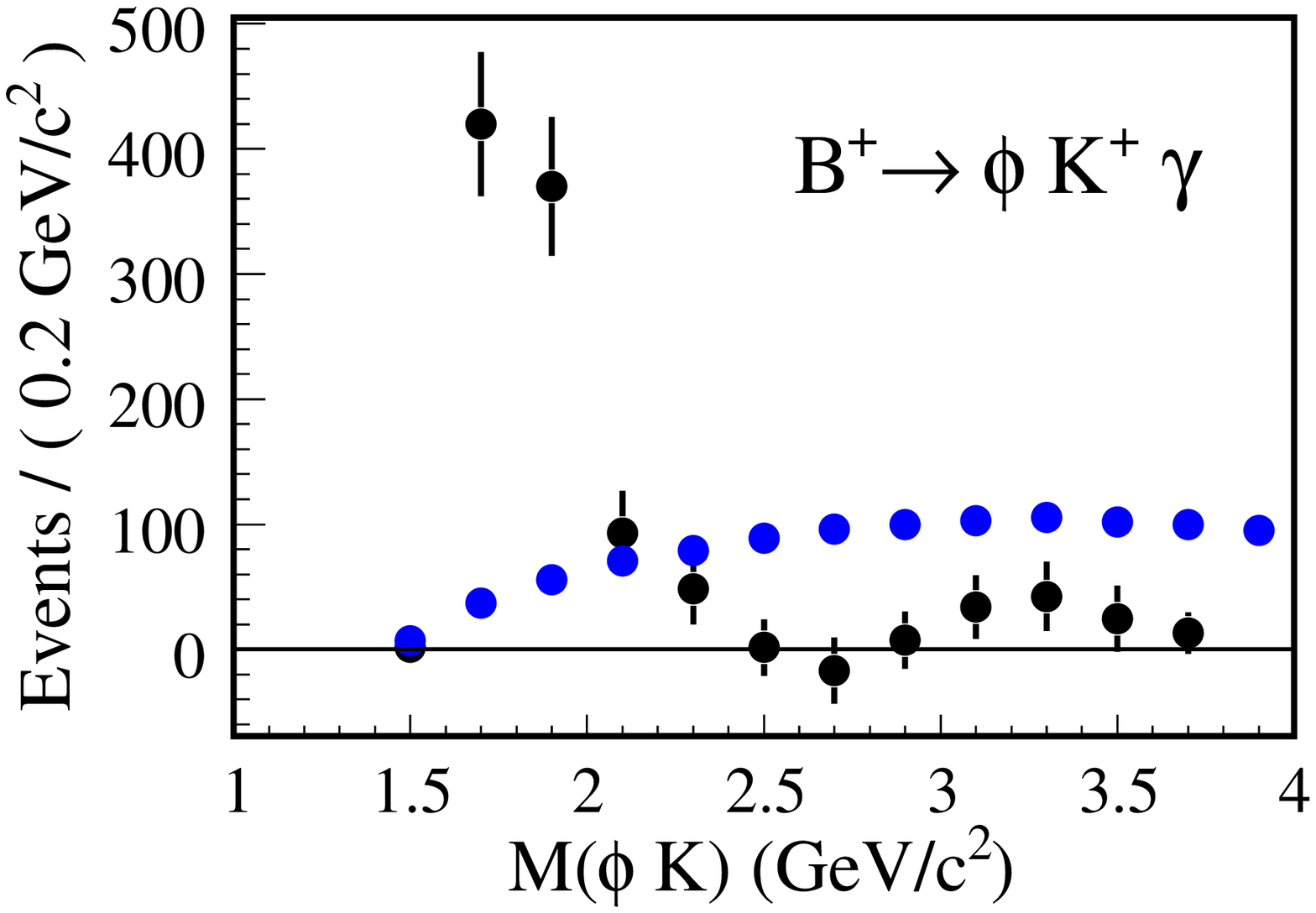}
    \includegraphics[width=6cm]{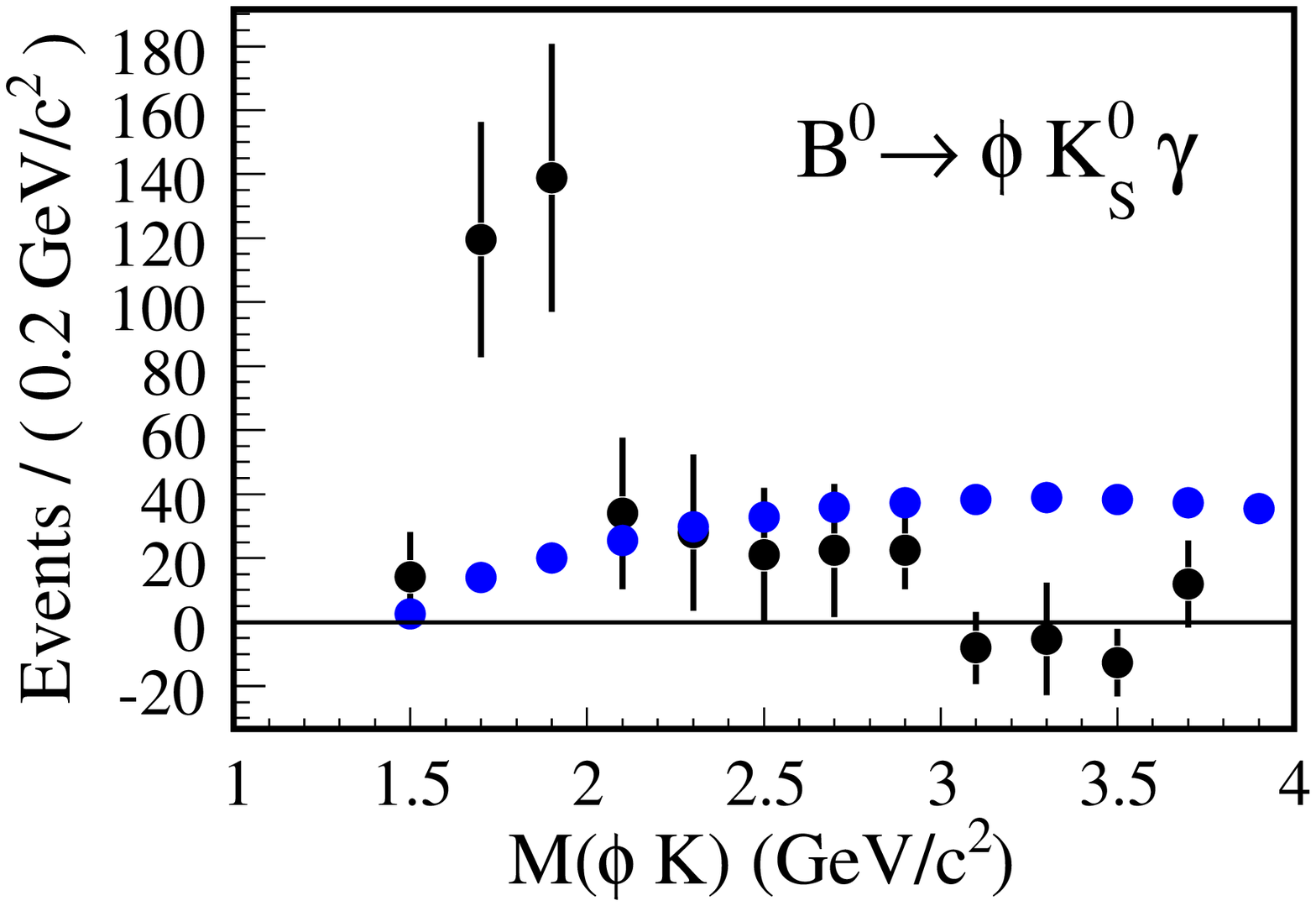}
    \caption{Background-subtracted and efficiency-corrected 
      $\phi K$ mass distributions for 
      the charged (left) and neutral (right) modes. 
      The points with error bars 
      represent the data. The yield in each bin is obtained by 
      the fitting procedure described in the text. 
      A three-body phase-space model from MC simulation 
      is shown by the filled circles (blue points) and 
      normalized to the total 
      data signal yield.}
    \label{fig:mxs}
  \end{center}
\end{figure}

To measure the $M_{\phi K}$ distribution,
we repeat the fit in bins of $\phi K$ mass and the resulting signal yields
are corrected for the detection efficiency.
Nearly $72\%$ of the signal events are concentrated
in the low-mass region,
$M_{\phi K} \in \left[1.5,2.0\right]$ GeV/$c^2$.
The background-subtracted and efficiency-corrected $M_{\phi K}$
distributions are shown in Fig.~\ref{fig:mxs}.
These spectra are consistent with the expectations from
the pQCD model for non-resonant 
$B \to \phi K \gamma$ decays~\cite{hsiangnan}.
The form factor effect produces the peak at the threshold.
With the present statistics no clear evidence is found for the existence
of a kaonic resonance decaying to $\phi K$.
The reconstruction efficiencies 
after reweighting according to this $M_{\phi K}$ dependence are
$(15.3\pm0.1 (\mathrm{stat}))\%$ for the charged mode and 
$(10.0\pm0.1 (\mathrm{stat}))\%$ for the neutral mode. 
The measured branching fractions are~\cite{sahoo-phikgamma}
\begin{eqnarray}
{\mathcal B}(B^+ \to \phi K^+ \gamma) = (2.48\pm 0.30 \pm 0.24) \times 10^{-6} \; {\rm and} \nonumber \\
{\mathcal B}(B^0 \to \phi K^0 \gamma) = (2.74\pm 0.60 \pm 0.32) \times 10^{-6}.
\end{eqnarray}

We also measure the charge asymmetry in 
$B^{\pm} \to \phi K^{\pm} \gamma$ decay to be~\cite{sahoo-phikgamma}
\begin{equation}
{\mathcal A}_{CP} = [N(B^-)-N(B^+)]/[N(B^-)+N(B^+)] 
= -0.03\pm 0.11\pm 0.08,
\end{equation}
where $N(B^-)$ and $N(B^+)$ are the signal yields 
for $B^-$ and $B^+$ decays, respectively.

We also performed the first time-dependent measurements in the
neutral $B^0 \to \phi K_S^0 \gamma$ mode.
In contrast to 
$B^0 \to K^{*0} (\to K_S^0 \pi^0) \gamma$, 
another related mode that is sensitive to NP, 
the time dependence of 
$B^0 \to \phi K_S^0 \gamma$ can be measured from the 
$\phi \to K^+K^-$ decay
and does not require a difficult measurement of the 
long lived $K_S^0$ decay inside the 
inner tracking volume or 
reconstruction of a low energy $\pi^0$.
The vertex position of the signal side is reconstructed using the
two kaon tracks from the $\phi$ meson.
Since the NR component is expected to have the same 
NP as the signal 
$B \to \phi K \gamma$, we treat this as signal for the
time-dependent fit.

The measured $CP$ violation parameters are~\cite{sahoo-phikgamma}
\begin{eqnarray}
{\mathcal S} = +0.74^{+0.72}_{-1.05} (\rm stat)^{+0.10}_{-0.24} (\rm syst) \; {\rm and} \nonumber \\
{\mathcal A} = +0.35 \pm 0.58 (\rm stat)^{+0.23}_{-0.10} (\rm syst).
\end{eqnarray}

We have established that the mode 
$B^0 \to \phi K_S^0 \gamma$ can be used
at future high luminosity $e^+e^-$
and hadronic facilities to perform
time-dependent $CP$ violation measurements and 
to carry out sensitive tests for NP.
The results are accepted for publication in Physical Review D~\cite{sahoo-phikgamma}.
\begin{figure}[htbp]
  \begin{center}
    \includegraphics[width=6cm]{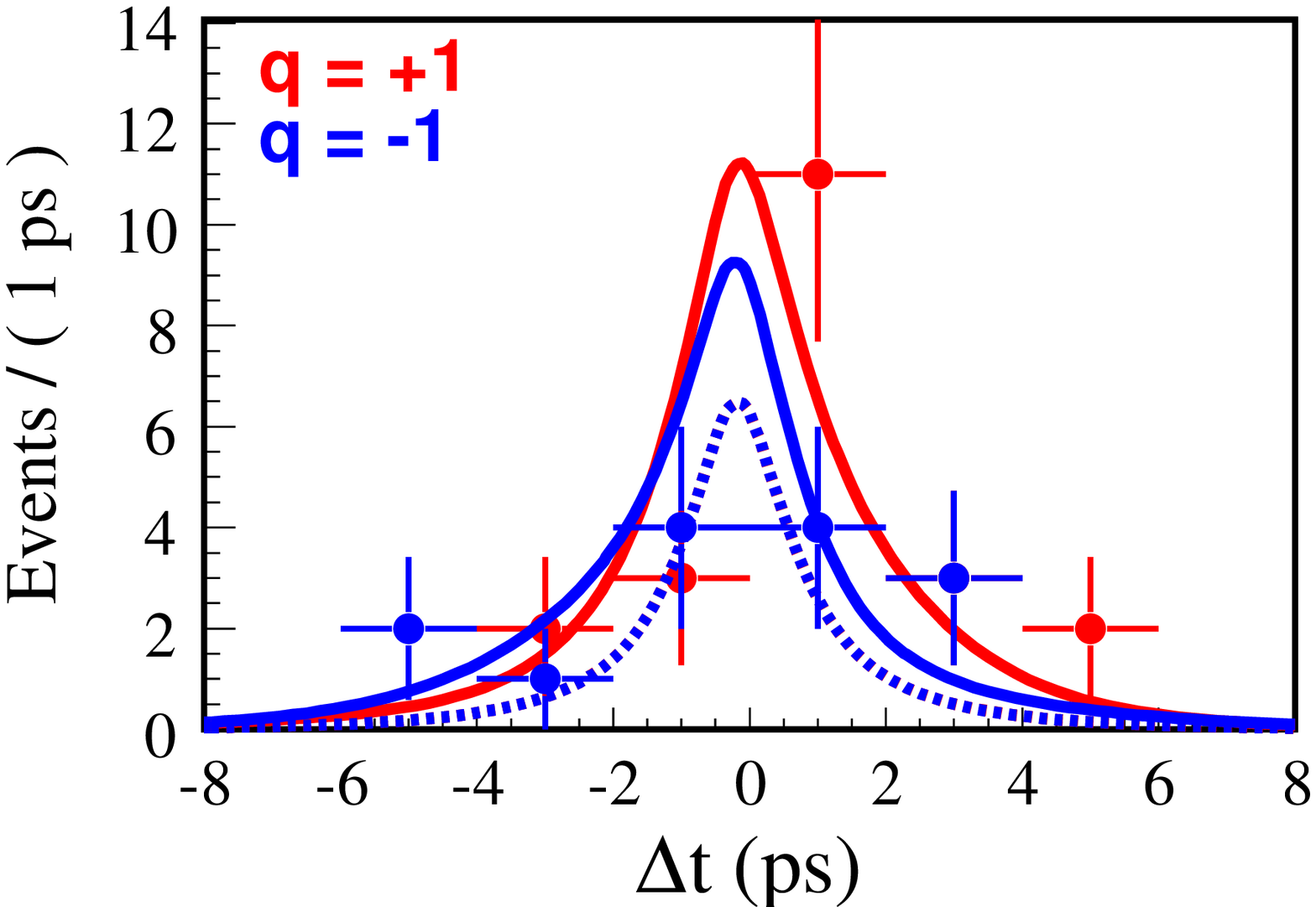}
    \includegraphics[width=6cm]{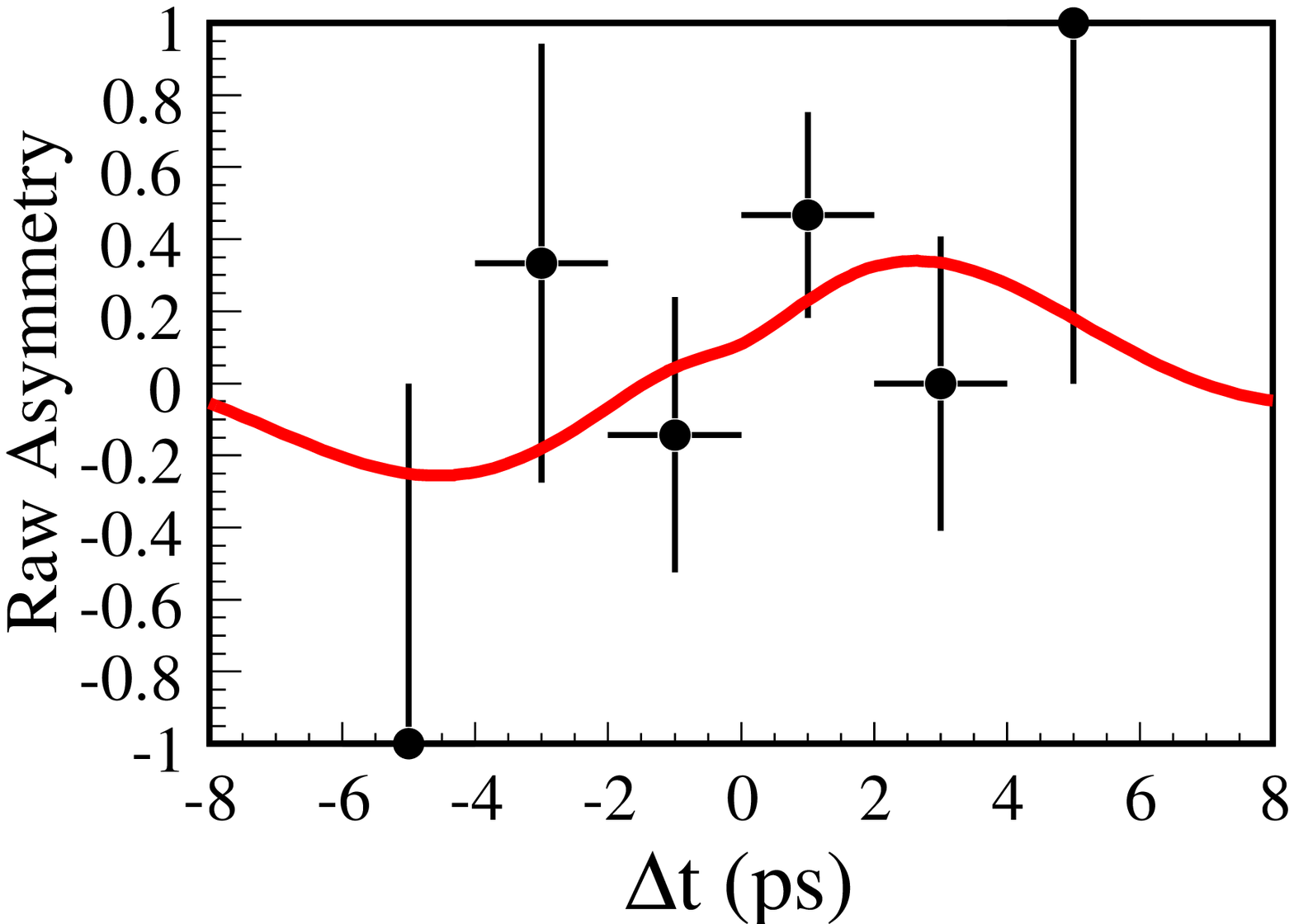}
    \caption{$\Delta t$ distributions for $q$ = $+1$ and $q$ = $-1$ (left)
      and the raw asymmetry (right) for well-tagged events. 
      The dashed curves in the $\Delta t$ plot are the sum of 
      backgrounds while the solid curves are the sum of signal and
      backgrounds. The solid curve in the asymmetry plot shows the result
      of the UML fit.}
    \label{fig:dtandasymm}
  \end{center}
\end{figure}

\begin{figure}[htbp]
\begin{center}
\includegraphics[width=5cm]{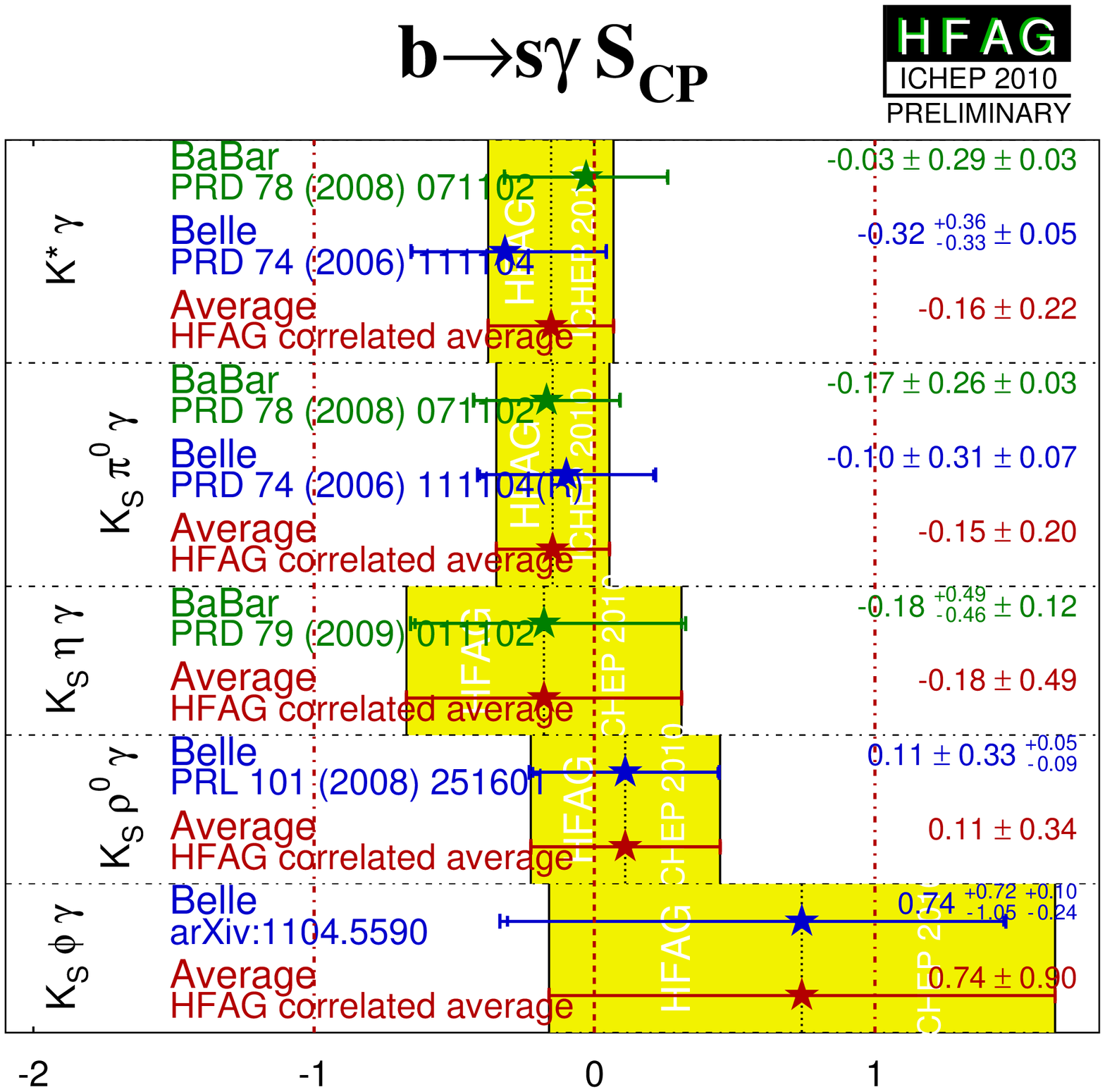}
\includegraphics[width=5cm]{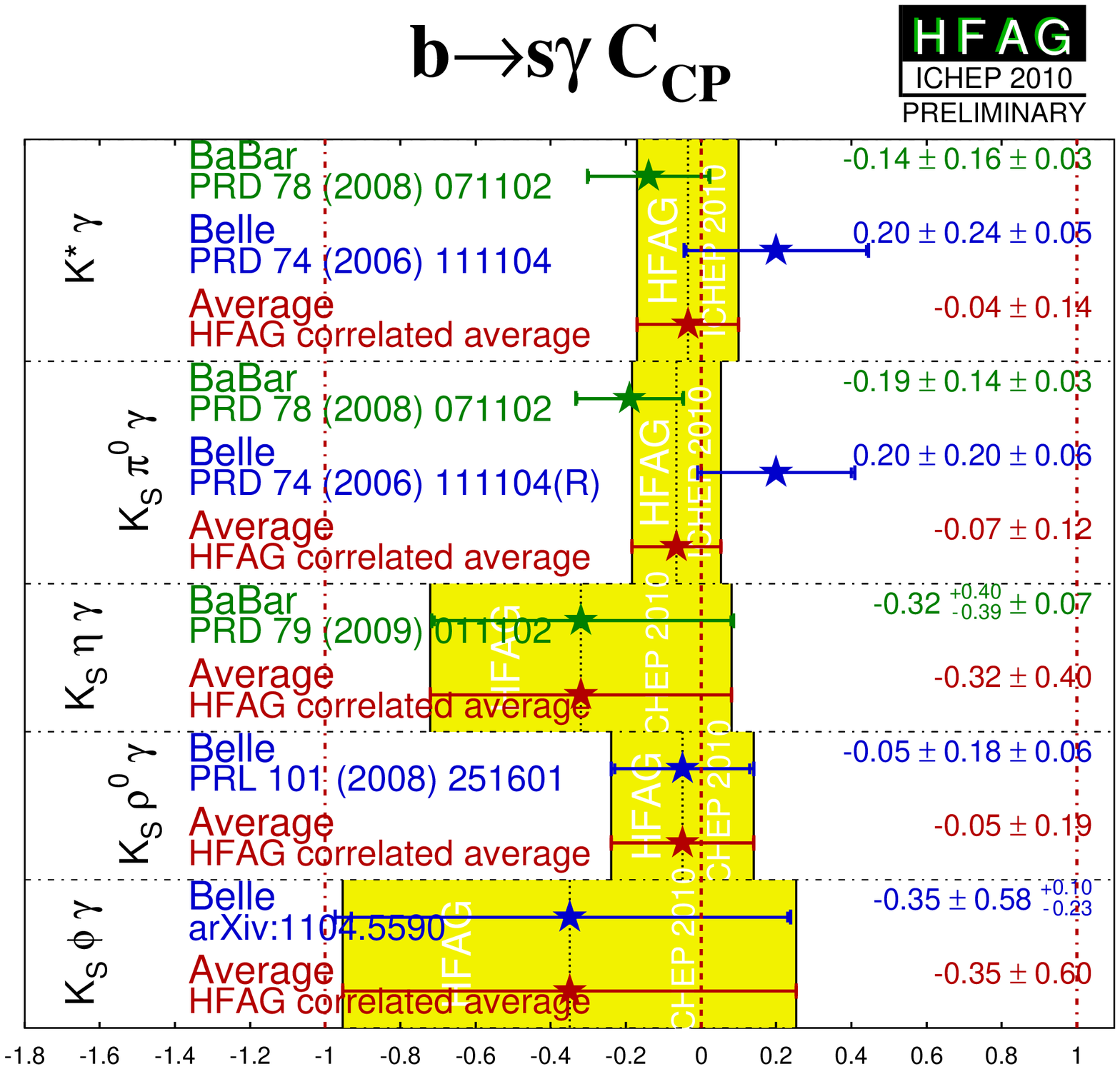}
\caption{The HFAG summary of the TCPV results in exclusive $b \to s \gamma$ decays from $B$ factories.}
\label{fig:hfag-btosgamma}
\end{center}
\end{figure}

The HFAG summary of the results in $b \to s \gamma$ modes from 
Belle and BaBar is shown in Fig.~\ref{fig:hfag-btosgamma}.
With the present statistics, these measurements are consistent
with the SM predictions and there is no indication of NP from 
right-handed currents in radiative $B$ decays.

\section{Summary}

In summary, we have presented the recent measurements on $CP$ violation
parameters from Belle using the full data sample 
$772 \times 10^6$ $B\bar{B}$ collected at the $\Upsilon(4S)$ resonance.
The $CP$ violation parameters in $b \to c\bar{c}s$ decays are the most
precise measurement and provide a reference point 
for new physics searches.
We have updated the branching fraction and $CP$ violation results
in $B^0 \to D^+D^-$ and $B^0 \to D^{*+}D^{*-}$ using the 
full data sample.  These are consistent
with the measurements of $\sin 2\phi_1$ from $b \to c\bar{c}s$ decays.
We also presented the first measurements of 
$CPT$ violation parameters in $B$ decays, which are consistent with zero.
The first observation of $B^0 \to \phi K^0 \gamma$ and its $CP$ violation
parameters are also reported. This mode establishes a new method for
searching right-handed currents and will be used in future very 
high luminosity experiments.

\begin{acknowledgments}
I would like to thank my Belle colleagues for giving me opportunity to present the
recent results on $CP$ violation measurements.
We thank the KEKB group for excellent operation of the
accelerator, the KEK cryogenics group for efficient solenoid
operations, and the KEK computer group and
the NII for valuable computing and SINET3 network support.  
We acknowledge support from MEXT, JSPS and Nagoya's TLPRC (Japan);
ARC and DIISR (Australia); NSFC (China); MSMT (Czechia);
DST (India); MEST, NRF, NSDC of KISTI, and WCU (Korea); MNiSW (Poland); 
MES and RFAAE (Russia); ARRS (Slovenia); SNSF (Switzerland); 
NSC and MOE (Taiwan); and DOE (USA).
\end{acknowledgments}

\bigskip 

\begin{thebibliography}{9}
\bibitem{conjugate}
Throughout this proceeding, the inclusion of the charge-conjugate
decay mode is implied unless otherwise stated.
\end{thebibliography}

\end{document}